\documentclass{aa}

\usepackage{graphicx}
\usepackage{amsmath}
\usepackage{txfonts}
\usepackage[T1]{fontenc}
\usepackage{booktabs}
\usepackage{multirow}
\usepackage{siunitx}
\usepackage{xcolor}
\usepackage[colorlinks,citecolor=blue,linkcolor=blue,urlcolor=blue]{hyperref}

\newcommand{\zl}{z_{\mathrm{l}}}
\newcommand{\zs}{z_{\mathrm{s}}}

\newcommand{\Farc}{F_{\mathrm{arc}}}

\newcommand{\SExtractor}{\texttt{SExtractor}}

\begin{document}

\title{The PAU Survey: Dissecting strong-lensing systems with narrow-band photometric redshifts}

\author{Shuaibo Geng\inst{1,2}
    \and Pablo Renard\inst{4,5}
    \and Qiuhan He\inst{7}
    \and David Navarro-Gironés\inst{6}
    \and Magherita Grespan\inst{1,8}
    \and Agnieszka Pollo\inst{1,3}
    \and Enrique Gazta\~naga\inst{4,9}
    \and Marek Biesiada\inst{1}
    \and Ricard Casas\inst{4,5}
    \and Juan Garc\'ia-Bellido\inst{10}
    \and Pau Tallada-Cresp\'{i}\inst{11,12}
    \and Jieci Wang\inst{2}
}

\institute{National Centre for Nuclear Research, ul. Pasteura 7, 02-093 Warsaw, Poland \\
\email{Shuaibo.Geng@ncbj.gov.pl}
    \and
    Hunan Normal University, 36 Lushan Road, Yuelu District, Changsha, Hunan 410081, China
    \and
    Astronomical Observatory of the Jagiellonian University, Faculty of Physics, Astronomy and Applied Computer Science, ul. Orla 171, 30-244 Kraków, Poland
    \and
    Institute of Space Sciences (ICE, CSIC), Campus UAB, Carrer de Can Magrans, s/n, 08193 Barcelona, Spain
    \and
    Institut d'Estudis Espacials de Catalunya (IEEC), 08860 Castelldefels (Barcelona), Spain
    \and
    Leiden Observatory, Leiden University, Einsteinweg 55, 2333 CC Leiden, The Netherlands
    \and
    Kapteyn Astronomical Institute, University of Groningen, Landleven 12, 9747 AD Groningen, The Netherlands
    \and
    Sub-department of Astrophysics, University of Oxford, Denys Wilkinson Building, Keble Road, Oxford OX1 3RH, UK
    \and
    Institute of Cosmology and Gravitation, University of Portsmouth, Dennis Sciama Building, Burnaby Road, Portsmouth PO1 3FX, UK
    \and
    Instituto de F\'isica Te\'orica CSIC/UAM, Universidad Aut\'onoma de Madrid, 28049 Madrid, Spain
    \and 
    Port d’Informaci\'o Cient\'ifica (PIC), Campus UAB, C. Albareda s/n, 08193 Bellaterra (Barcelona), Spain
    \and
    Centro de Investigaciones Energéticas, Medioambientales y Tecnológicas (CIEMAT), Avenida Complutense 40, 28040 Madrid, Spain
}

\abstract
{Wide-field imaging surveys, like Euclid, will identify large samples of strong-lens candidates. Their scientific use often requires redshift information for both the deflector and the lensed source. Source redshifts are especially difficult to obtain because the arcs are faint and contaminated by foreground light.}
{We present a proof-of-concept demonstration using eight known strong-lens systems to assess agreement with spectroscopic redshifts and evaluate the contributions of narrow-band photometry and foreground-light subtraction. The resulting redshift information can help validate lens candidates by testing the lens-source redshift ordering and provide accurate source redshifts in favourable cases.}
{Regions of different components are difined by reference broad-band lens models. We subtract a point-reflected region to reduce foreground contamination and combine the retained PAU Survey narrow bands with five HSC or CFHT broad bands. BCNz then estimates separate redshift distributions for the lens and source, without imposing their expected ordering.}
{All eight lens peaks agree with spectroscopy within $|\delta_z|\leq0.014$, where $\delta_z=(z_b-z_{\rm spec})/(1+z_{\rm spec})$. Four source peaks, all at $z_{\rm s}^{\rm spec}\leq1.23$, have $|\delta_z|\leq0.025$. Symmetric subtraction lowers the median source $|\delta_z|$ from 0.336 to 0.071. At the common descriptive threshold $|\delta_z|<0.05$, PAUS NB+BB yields eight lens and four source matches, compared with three and two for EAZY using broad bands alone.}
{Our results show that lens-model-informed narrow-band photometry can recover component redshift information even when source light is blended with the deflector. It motivates further development of redshift constraints as a complement to morphology and colour in lens assessment. Wider narrow-band surveys such as J-PAS offer opportunities to extend the approach to larger samples and support follow-up prioritisation, with uncertainty calibration and blind candidate tests as the next steps.}


\maketitle

\section{Introduction}
\label{sec:intro}

Strong gravitational lensing is sensitive to both the mass distribution of the lens and the angular-diameter distances between the observer, lens, and source, making it a powerful probe of galaxy structure and cosmology. Its astrophysical applications include constraining the total mass distributions of galaxies, encompassing both dark and luminous matter \citep[e.g.][]{sonnenfeld2019}, investigating magnified high-redshift sources \citep[e.g.][]{vanderwel2013}, detecting dark matter subhaloes \citep[e.g.][]{vegetti2012,he2023subhalo,nightingale2024}, and placing statistical constraints on low-mass halo populations \citep[e.g.][]{gilman2020,he2022}. Cosmological applications range from measuring $H_0$ through time-delay cosmography \citep{tdcosmo2025} and constraining dark energy using double-source-plane lenses \citep{collettAuger2014} to testing cosmological models through lensing distance ratios combined with stellar kinematics \citep{cao2012cosmology,cao2015cosmology}. Accurate and reliable redshift measurements for both lenses and sources are crucial for these applications and for fully exploiting the rapidly growing samples of strong gravitational lenses.

Wide-field strong-lens searches combine automated candidate selection with visual grading and, where available, spectroscopic validation. In HSC-SSP, SuGOHI progressed from YattaLens, which combines lens-light subtraction, arc detection, colour selection, and lens modelling \citep{sonnenfeld2018,wong2022}, to convolutional neural networks \citep{jaelani2024}. The latter approach identified 43 Grade-A and 269 Grade-B candidates in SuGOHI-X after visual inspection \citep{jaelani2024}. KiDS searches likewise adopted neural networks \citep{petrillo2017} and subsequently transformer architectures through TEGLIE \citep{grespan2024}. Transformers use self-attention to capture spatial relationships across images, complementing the architectures used in earlier searches \citep{vaswani2017,dosovitskiy2021}. Recent DES Year-6 and Euclid Q1 searches demonstrate the application of vision transformers to large imaging samples \citep{gonzalez2026,vincken2026}. The DES search alone examined approximately 236 million image cutouts \citep{gonzalez2026}. These methods have greatly expanded candidate catalogues, increasing the need for efficient physical validation.

Expert inspection remains valuable for recognising complex lensing configurations and rejecting contaminants that are difficult to represent in automated training sets. However, its selection function is difficult to quantify and exhaustive grading becomes costly as samples grow. Moreover, morphology and colour provide little direct information on the relative redshifts of the deflector and candidate source. Only a small fraction of high-ranking candidates currently have spectroscopic validation \citep{sonnenfeld2018,wong2022}, leaving a gap between efficient image-based discovery and secure physical confirmation.

Spectroscopic follow-up provides the most direct way to confirm the redshifts of strong-lens systems, but it is observationally expensive and cannot be obtained for every candidate expected from future wide-field surveys, such as Vera C. Rubin Observatory Legacy Survey of Space and Time (LSST). The Dark Energy Spectroscopic Instrument (DESI) is a highly multiplexed spectroscopic facility capable of observing up to 5,000 targets simultaneously using robotically positioned optical fibres \citep{desi2022}. However, even when DESI successfully measures the lens redshift, foreground-light contamination, limited signal-to-noise ratio, and wavelength coverage can still prevent the recovery of the background source redshift \citep{huang2025desi}. 

Photometric redshifts provide complementary information and already support precision cosmology: DES Year 6 combined multi-band photometry with clustering-redshift measurements to calibrate its weak-lensing source redshift distributions \citep{yin2025}. For strong lenses, combining broad- and narrow-band imaging offers additional spectral information. The PAU Survey (PAUS) uses 40 narrow bands spanning approximately $4500$--$8500$~\AA\ to sample spectral energy distributions as low-resolution spectra \citep{padilla2019,eriksen2019}. The narrow bands localise continuum breaks and emission-line features, while broad bands constrain the overall continuum shape. Compared with broad-band colours alone, this dense wavelength sampling helps distinguish spectral features that would otherwise be smeared across a filter, while retaining the scalability of imaging surveys. In COSMOS, this combination achieved $\sigma_{68}/(1+z)=0.0037$ for the best-measured half of galaxies with $i_{\rm AB}<22.5$ \citep{eriksen2019}, with further improvements from ancillary photometry and refined modelling \citep{alarcon2021}. This spectral information, obtained through imaging, makes PAUS an attractive test bed for redshift-based assessment of strong-lens candidates.

Bright, red early-type galaxies dominate the deflector population and can usually be handled well by photometric-redshift pipelines. In contrast, the lensed source is relatively faint and is not isolated from the deflector light.
By construction it appears around the Einstein radius, typically $1$--$2$~arcsec from the deflector centre, which in a seeing-limited ground-based image is one to two full widths at half maximum. The lensed source image is superposed on the deflector's envelope, which at that radius is both bright and steeply falling. A circular or polygonal region placed on the arc measures the sum of the two. Extracting the source colours through lens modelling can still have 75\% photo-z outlier fraction when lens surface brightness is higher than the source at the image position \citep{langeroodi2023}.
However, early-type galaxies are among the most morphologically symmetric galaxy populations. Their point asymmetries are substantially lower than those of spiral, irregular, and star-forming galaxies \citep{conselice2000,conselice2003}.
This symmetry can be exploited to remove the deflector contamination without any assumption about the source's own spectral energy distribution.

In this work, we present a proof-of-concept demonstration of lens-model-informed photometry that combines broad- and narrow-band imaging to estimate the photometric redshifts of lens and source galaxies, respectively, for eight selected strong-lensing systems. We compare the inferred redshift distributions of both components with spectroscopic measurements and conduct matched photometric control tests to quantify the contributions of narrow-band photometry and point-reflection subtraction. Our study explores how component-resolved redshift information can complement morphological and colour-based diagnostics in the assessment of strong-lens systems.

The paper is structured as follows. Section~\ref{sec:sample} describes the HSC-SSP, CFHTLS, and PAUS imaging and the selection of the eight analysed lens systems. Section~\ref{sec:method} presents the method: broad-band lens modelling, image-based selection of the lens, source, and reflected-background regions, reflected-pair photometry, and the BCNz redshift fits. Section~\ref{sec:results} presents the redshift recovery, the ordering test, and matched control fits without symmetric subtraction or narrow bands. Section~\ref{sec:discussion} discusses the conditions under which both redshifts are recovered, the role of the method in lens-candidate validation, its limitations, and prospects for future surveys and applications beyond strong lensing. Section~\ref{sec:conclusions} summarises our conclusions. Appendix~\ref{app:other-systems} shows the results for the remaining four systems, and Appendix~\ref{app:setup} lists the reference-imaging and reproducibility details.

\section{Data and sample}
\label{sec:sample}
In this section, we describe the narrow-band and broad-band imaging data used
in this work, together with the selected strong-lens systems with confirmed
spectroscopic redshifts.

\begin{table*}
\caption{Strong-lens systems considered for the narrow-band analysis.
The spectroscopic lens and source redshifts retain the precision
reported in the original references.}
\label{tab:strong-lens-sample}
\centering
\small
\setlength{\tabcolsep}{4pt}
\begin{tabular}{lrrcccccc}
\toprule
Lens name & RA & Dec & $z_{\rm l}^{\rm spec}$ & $z_{\rm s}^{\rm spec}$ &
NB completeness & HSC field & CFHT field & \shortstack{Photometric\\analysis} \\
\midrule
SL2SJ0214$-$0405       &  33.54672 & $-4.08414$ & 0.61 & 1.88 & 30/40 & in  & in & No\\
SL2SJ0217$-$0513       &  34.40513 & $-5.22475$ & 0.64616 & 1.85 & 40/40 & in  & in & Yes \\
J0223$-$0534           &  35.94230 & $-5.57190$ & 0.499 & 1.444 & 40/40 & in  & in & Yes \\
J0225$-$0454           &  36.29600 & $-4.90932$ & 0.23795 & 1.2 & 40/40 & in  & in & Yes \\
J0226$-$0420           &  36.54450 & $-4.33660$ & 0.494 & 1.23 & 40/40 & in  & in & Yes \\
J0907$+$0003           & 136.79370 & $ 0.05590$ & --   & 1.30 & 40/40 & in  & out & No\\
J0907$-$0042         & 136.91667 & $-0.70000$ & 0.61 & 1.58 & 40/40 & in  & out & No\\
J0910$-$0003         & 137.67917 & $-0.05611$ & 0.79 & 1.79 & 40/40 & in  & out & No\\
SDSSJ0912$+$0029  & 138.02212 & $ 0.48364$ & 0.1642 & 0.3239 & 26/40 & in  & out & Yes \\
J1404$+$5200           & 211.22692 & $52.00686$ & 0.456 & 1.59 & 40/40 & out & in & Yes \\
J1406$+$5226           & 211.70970 & $52.43870$ & 0.716 & 1.47 & 40/40 & out\tablefootmark{*}  & in & Yes \\
SDSSJ1416$+$5136  & 214.09304 & $51.60839$ & 0.2987 & 0.8115 & 30/40 & out\tablefootmark{**} & in & Yes \\
HSTJ1418$+$5236            & 214.58654 & $52.60314$ & 0.58 & 0.98 & 40/40 & in  & in & No\\
J1420$+$5231           & 215.01420 & $52.52720$ & 0.354 & 1.41 & 40/40 & in  & in & No\\
\bottomrule
\end{tabular}
\tablefoot{RA and Dec are given in degrees.
NB completeness is the number of available PAUS narrow bands out of 40. The
HSC and CFHT columns indicate whether usable broad-band imaging products are
available for the analysis. A dash denotes an unavailable value.
Photometric analysis indicates whether a system is included in the component photometry and photo-$z$ results presented in this work: Yes for the eight analysed systems and No for the remaining six.
\tablefoottext{*}{No usable HSC image was available for J1406$+$5226.}
\tablefoottext{**}{HSC imaging was obtained for SDSSJ1416$+$5136,
but no usable HSC PSF model was available.}}
\end{table*}

\subsection{Broad-band imaging}
\label{sec:bb imaging}

We use broad-band imaging from the Hyper Suprime-Cam Subaru Strategic Program (HSC-SSP) and the Canada-France-Hawaii Telescope Legacy Survey (CFHTLS).

The CFHTLS is a wide-field optical imaging survey carried out with MegaCam on the 3.6~m Canada-France-Hawaii Telescope (CFHT) \citep{gwyn2012}. Its Wide component covers approximately 150~deg$^2$ across four fields and provides imaging in five broad bands ($ugriz$). The final stacked images reach typical point-source depths of $u\simeq26.0$, $g\simeq26.5$, $r\simeq25.9$, $i\simeq25.7$, and $z\simeq24.6$~mag \citep{gwyn2012}.

The HSC-SSP is conducted with Hyper Suprime-Cam (HSC) on the 8.2~m Subaru Telescope \citep{aihara2018,miyazaki2018}. HSC provides a 1.5-degree-diameter field of view, enabling efficient wide-area imaging with an 8--10~m class telescope \citep{miyazaki2018}. The HSC-SSP Wide layer covers approximately 1\,400~deg$^2$ in five broad bands ($grizy$), with a target depth of $i\sim26$~mag and typical image quality of $\sim0.6$~arcsec in the $i$ band \citep{aihara2018}.

Both surveys provide deep, multi-band imaging with sub-arcsecond spatial resolution, enabling detailed modelling of foreground lens light and reliable identification of lensed features for subsequent component photometry. The HSC-SSP also provides a useful precursor data set for the LSST, which will likewise obtain deep multi-band optical imaging over a much larger sky area \citep{ivezic2019}. Although the two surveys differ in telescope aperture, survey area, cadence, and image quality, many analysis challenges relevant to strong-lens discovery and modelling in LSST can already be studied with HSC data. Both CFHTLS and HSC-SSP therefore provide the spatial information needed to model the foreground lens light and to define the arc regions used in the subsequent narrow-band photometric analysis.

\subsection{Narrow-band imaging}
\label{sec:nb imaging}

The narrow-band imaging is provided by the Physics of the Accelerating Universe Survey (PAUS), carried out with the Physics of the Accelerating Universe Camera (PAUCam) at the prime focus of the 4.2~m William Herschel Telescope (WHT) on La Palma \citep{padilla2019,marti2014,serrano2023}. PAUS covers several tens of square degrees, primarily across the CFHTLS W1 and W3 fields, the GAMA G09 field overlapping with KiDS (often referred to as W2 in PAUS studies), and the Cosmic Evolution Survey (COSMOS) field, with additional limited coverage of CFHTLS W4.

PAUCam is equipped with 40 narrow-band filters uniformly distributed between 455 and 845~nm in steps of 10~nm. Each filter has a full width at half maximum of approximately 12.5~nm, with about 25 per cent overlap between adjacent bands \citep{marti2014,serrano2023}. This configuration provides a spectral resolution of approximately $R\simeq50$. This is relatively low compared with conventional spectroscopy but substantially higher than that of standard broad-band photometry. The dense wavelength sampling allows PAUS to localise continuum breaks and emission-line features that are only roughly sampled by broad-band filters \citep{marti2014,eriksen2019}. This capability has enabled measurements of the 4000~\AA\ break \citep{renard2022d4000} and the selection of quasars through Ly$\alpha$ emission \citep{torralba2024lya}.

For ordinary field galaxies, PAUS has demonstrated quasi-spectroscopic
photometric-redshift performance. Using the narrow-band data together with
broad-band photometry in the Cosmic Evolution Survey field, PAUS reaches
$\sigma_{68}/(1+z)=0.0037$ for the best-measured half of galaxies with
$i_{\rm AB}<22.5$ over $0<z<1.2$ \citep{eriksen2019}. For bright,
high-quality objects, the precision can approach $\sim0.001$, mainly because the narrow bands allow prominent emission-line features to be identified \citep{eriksen2019}. Subsequent work further improved the PAUS photometric redshifts by combining the narrow-band measurements with extensive ancillary photometry and refined spectral-energy-distribution modelling \citep{alarcon2021}. This combination of spectral information and imaging scalability makes PAUS particularly suitable for testing whether narrow-band photometry can recover independent redshift information for the deflector and source components of strong-lens systems.

\subsection{Strong lensing sample}
\label{sec:strong lensing sample}

We constructed the parent sample of galaxy-scale strong lensing systems by cross-matching the PAUS footprint with strong-lens systems reported in the literature. The resulting catalogue contains 58 systems drawn from several surveys. The largest contribution is 30 systems from the Strong Lensing Legacy
Survey (SL2S) \citep{cabanac2007,more2012,sonnenfeld2013,gavazzi2014}. A further 18 systems were discovered in Hubble Space Telescope (HST) imaging. These comprise eight systems from the crowdsourced Hubble Asteroid Hunter search \citep{garvin2022}, five from the COSMOS or CANDELS fields
\citep{jackson2008,vanderwel2013,pourrahmani2018}, three from earlier HST Medium Deep Survey or archival WFPC2 searches \citep{fischer1998,ratnatunga1995,sand2005}, and two from the HST/ACS AEGIS survey \citep{moustakas2007}. Four submillimetre-selected lenses from the Herschel Astrophysical Terahertz Large Area Survey (H-ATLAS) Science Demonstration Phase \citep{negrello2010}. Three were selected from Sloan
Digital Sky Survey spectroscopy: two from the SDSS/SLACS lens search and one from the SDSS-III/BOSS quasar-lens survey \citep{bolton2004,more2016}. Two
were identified or confirmed in HSC-SSP imaging by SuGOHI \citep{chan2020,jaelani2021}. Finally, J0907$+$0003 was first reported by the KiDS-SQuaD search \citep{spiniello2018} and was later identified
independently in Gaia and HSC searches \citep{kronemartins2019,chan2020}.

Fourteen systems from the parent catalogue lie within the HSC-SSP or CFHTLS footprints and have spectroscopic source redshifts below 2 (Table~\ref{tab:strong-lens-sample}). This limit reflects the $0.01\leq z\leq2.0$ range used in our PAUS narrow-band photo-$z$ analysis, consistent with previous PAUS wide-field work \citep{navarroGirones2024}. The spectroscopic redshifts of these systems define the sample selection and serve as reference values for assessing the photo-z estimates (Sect.~\ref{sec:results}). They do not enter the photo-$z$ fits as priors or input values. We compiled the redshifts from the SL2S and SuGOHI studies \citep{more2012,sonnenfeld2013,sonnenfeld2019}, the KiDS-SQuaD and SuGOHI analyses \citep{spiniello2018,chan2020}, the H-ATLAS studies \citep{negrello2010,dye2014,wong2017}, the SLACS survey \citep{bolton2004,bolton2008}, and the HST/ACS AEGIS survey \citep{moustakas2007}.

We attempted the broad-band lens modelling described in Sect.~\ref{sec:method} for all 14 systems. Four systems (SL2SJ0214$-$0405, J0907$-$0042, J0910$-$0003, and HSTJ1418$+$5236) did not yield robust models. Their small image separations, corresponding to characteristic Einstein scales below approximately 1~arcsec, are difficult to resolve in the available ground-based broad-band images. J0907$+$0003 is a lensed quasar system dominated by bright source emission, and its lens spectroscopic redshift is unavailable. Our method targets extended arcs blended with deflector light, so this system lies outside the scope of the present analysis.
J1420$+$5231 is also excluded from the analysis because its deflector is a disk galaxy. Our reflected-region subtraction assumes approximately point-symmetric deflector light (Sect.~\ref{sec:photometry}), which is less appropriate for a disk lens. We therefore list this system in Table~\ref{tab:strong-lens-sample} but do not include it in the results. The results below therefore refer to the other eight systems.

Three systems have measurements in fewer than the full set of 40 PAUS narrow bands (Table~\ref{tab:strong-lens-sample}). This incomplete coverage mainly arises from the PAUCam focal-plane layout and dither pattern, together with the rejection of exposures affected by detector gaps, vignetting, bad pixels, cosmic rays, poor seeing, or failed quality checks \citep{serrano2023}. Missing narrow bands do not affect our lens modelling, which relies on HSC-SSP or CFHTLS broad-band images, but may reduce the information available for photometric-redshift estimation.

Previous PAUS studies have investigated the impact of incomplete narrow-band coverage. \citet{eriksen2020} showed that photo-$z$ performance can deteriorate when too few exposures or bands are retained. However, \citet{navarroGirones2024} found negligible differences in the overall $\sigma_{68}$ when comparing field-galaxy samples requiring at least 30 narrow bands and all 40 bands. These findings suggest that moderate incompleteness does not necessarily compromise the statistical photo-$z$ performance of a galaxy sample. For individual strong-lens components, however, the impact depends on which bands are missing. In particular, the absence of bands covering prominent spectral breaks or emission lines may weaken redshift constraints and produce broader or multimodal posteriors.

\section{Method}
\label{sec:method}

\begin{figure*}[h]
\centering
\includegraphics[width=\textwidth]{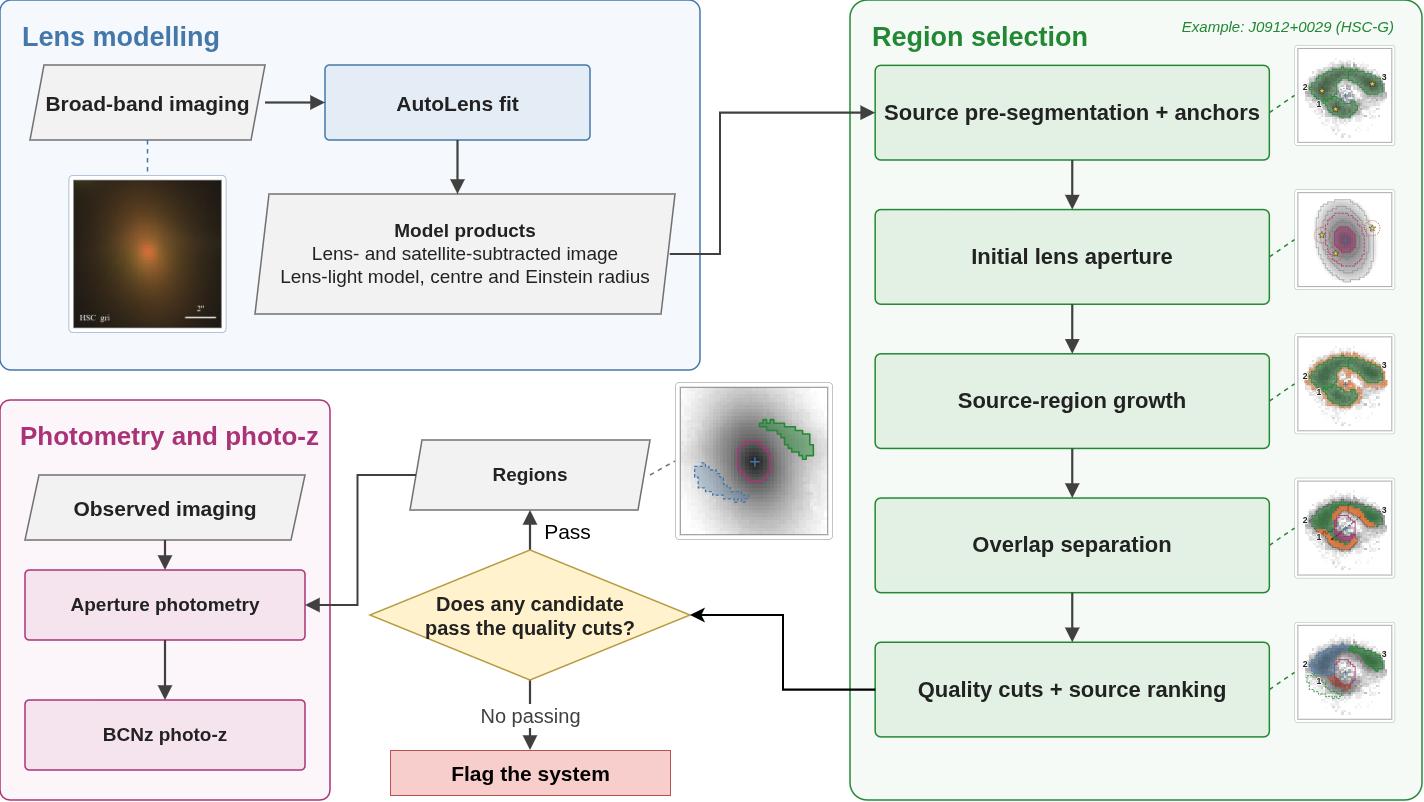}
\caption{Workflow of the strong-lensing photometric-redshift estimation pipeline developed in this work. Broad-band lens modelling provides the source-detection image and reference geometry. The right-hand block illustrates the successive region-selection steps, in which source candidates are grown, cleaned, and evaluated through quality cuts and ranking. The lens region is defined using an MGE isophote and subsequently refined to preserve symmetry. The resulting regions are used for component photometry on the observed images, as shown in the lower-left block. Finally, the extracted lens and source photometry is separately processed with BCNz to estimate their respective redshift probability distributions.
}
\label{fig:workflow}
\end{figure*}

As shown in Fig.~\ref{fig:workflow}, the pipeline has three main steps:
broad-band lens modelling, image-based region selection, and multi-band photometry with redshift estimation.
The narrow-band images have low angular resolution, so we use the broad-band images to model the lens system and define the regions. The lens model provides the reference geometry and a lens-subtracted image for source-region selection.
We then apply the resulting regions to the observed narrow-band images and measure the fluxes directly from the data. The measured fluxes come from the images, not from the fitted model amplitudes.

\subsection{Broad-band lens modelling}
\label{sec:model}

We fit one reference broad-band image per system with \texttt{PyAutoLens} \citep{nightingale2021}. The deflector light is represented by a multi-Gaussian expansion \citep[MGE;][]{emsellem1994,cappellari2002,he2024},
\begin{equation}
 L(\boldsymbol{x})=\sum_k A_k
 G_k(\boldsymbol{x}-\boldsymbol{c};\sigma_k,q_k,\phi_k).
 \label{eq:mge}
\end{equation}
Here, $\boldsymbol{x}$ denotes the image-plane coordinates, $\boldsymbol{c}$ is the common centre, and $A_k$ is the amplitude of the $k$-th Gaussian component. Each Gaussian profile $G_k$ is characterised by its width $\sigma_k$, axis ratio $q_k$, and position angle $\phi_k$. We use two sets of 15 Gaussians with logarithmically spaced widths. In this work, the Gaussians within each set share an ellipticity, and both sets share the centre $\boldsymbol{c}$. Their amplitudes are determined using a non-negative linear solver. Modelled satellites are assigned separate light components. To facilitate reliable region selection through lens modelling, we adopt an elliptical power-law mass profile \citep{tessore2015} with external shear \citep{keeton1997} and reconstruct the source on a regularised pixel grid.

\texttt{PyAutoLens} requires lens and source redshift inputs, for which we adopt the common fiducial values $z_{\rm l}=0.5$ and $z_{\rm s}=1.0$ for every system. For the angular image modelling used here, these values affect the physical mass normalisation rather than the fitted image geometry. We verified that replacing the system-specific redshifts with this fiducial pair leaves the model images and selected regions unchanged. The resulting aperture photometry and BCNz redshift estimates are therefore unaffected.

At this stage, we export three products: the fitted light centre, the
lens-light model, and the exact lens-plane-subtracted image from the fit. The fitted light centre is used to define the point reflection for background estimation, while the output image subtracts the light of both the deflector and the modelled
satellite light components while preserving the lensed source.
Detecting and segmenting arcs in the original blended image can be
challenging because they are often faint against the foreground lens light.
This contamination can also bias source colours and photometric redshifts
\citep{sonnenfeld2018,langeroodi2023}.
We therefore perform segmentation and region selection on the
lens-plane-subtracted image, which provides a cleaner basis for source
detection. We describe this procedure in Sect.~\ref{sec:regions}.
The exported products retain the native image coordinates and WCS.

\subsection{Image-only region selection}
\label{sec:regions}

As shown in Fig.~\ref{fig:workflow}, region selection involves several
steps. First, we apply a single $4\sigma$ \SExtractor{} segmentation to the
lens-plane-subtracted image to identify source candidates
\citep{bertin1996}. Each retained component defines a candidate, and its
peak position serves as an anchor when setting the lens region. We exclude
small components, duplicate peaks, and peaks within half a PSF FWHM of the
lens centre from the anchor list.

Next, we define the initial lens region using an MGE isophote near the
fitted Einstein radius. We choose the isophote so that the region does not
enclose any valid source anchor. To capture more source flux, we then grow
each candidate from its $4\sigma$ seed to include connected emission down to
$2\sigma$.

The grown source regions may overlap the initial lens region. To minimise cross-contamination between the lens and source photometry, overlapping pixels should be assigned to the appropriate component. We assign overlapping pixels using an estimated source-light fraction, defined as
\begin{equation}
 f_{\rm s}(\boldsymbol{x})=
 \frac{\Delta I(\boldsymbol{x})}
 {L(\boldsymbol{x})+\Delta I(\boldsymbol{x})},
 \label{eq:point-fraction}
\end{equation}
where $\boldsymbol{x}$ denotes the image-plane coordinates,
$L(\boldsymbol{x})$ is the modelled lens-light intensity, and
$\Delta I(\boldsymbol{x})=I(\boldsymbol{x})-I(P\boldsymbol{x})$
is the residual intensity obtained from point-reflection subtraction.
Here, $I(\boldsymbol{x})$ denotes the observed image intensity,
and $P\boldsymbol{x}=2\boldsymbol{c}-\boldsymbol{x}$ represents
the reflection of $\boldsymbol{x}$ about the fitted lens centre
$\boldsymbol{c}$ in image coordinates.

An overlapping pixel is assigned to the source if
$0.4\leq f_{\rm s}\leq1$ and the denominator exceeds the
adopted noise threshold.
We exclude pixels overlapping other detected components or
modelled satellites, as well as pixels whose reflections
overlap these contaminants or the candidate itself.

Finally, we apply quality cuts to the surviving connected regions. Each
region must contain at least 20 pixels, have an integrated signal-to-noise
ratio (S/N) of at least 15, and retain at least 70 per cent
of its pre-carving S/N. Among candidates that pass these cuts, we select the
one with the highest effective S/N:
\begin{equation}
 ({\rm S/N})_{\rm eff}=
 \frac{F_{\rm diff}}
 {\sqrt{\sigma_{\rm stat}^{2}+(0.015\,|F_{\rm lens}|)^{2}}}.
 \label{eq:effective-snr}
\end{equation}
Here, $F_{\rm diff}$ is the candidate-minus-reflection flux in the
detection image, $\sigma_{\rm stat}$ is its statistical uncertainty,
and $F_{\rm lens}$ is the modelled lens flux within the candidate.
The lens-light term penalises candidates that are more susceptible
to lens-light leakage. It affects candidate ranking only and does
not contribute to the final photometric uncertainty.

The final background region is the point reflection of the selected source
region. Source-assigned pixels and their reflections are removed from the
lens region to preserve its symmetry. The regions are exported as WCS
polygons along pixel-cell boundaries and used directly for photometry.
A system is flagged if no candidate passes.

\subsection{Reflected-pair photometry}
\label{sec:photometry}

Astrometric residuals, differential atmospheric refraction and PSF variations can displace the lens centre by different amounts in each band. We therefore estimate a symmetry centre independently on each PAUS narrow-band coadd before photometry. For each narrow-band coadd, we take as the lens centre the point about which
the lens light is most nearly point-symmetric. This symmetry centre
$\boldsymbol{c}$ minimises
\begin{equation}
 Q(\boldsymbol{c})=\frac{1}{N_{\mathcal A}}
 \sum_{\boldsymbol{x}\in\mathcal{A}(\boldsymbol{c})}
 \frac{[I(\boldsymbol{x})-I(2\boldsymbol{c}-\boldsymbol{x})]^2}
 {V(\boldsymbol{x})+V(2\boldsymbol{c}-\boldsymbol{x})},
 \label{eq:chi2}
\end{equation}
where $I$ and $V$ are the pixel intensity and variance, and the sum runs over the $N_{\mathcal A}$ valid pixels of an annulus $\mathcal{A}(\boldsymbol{c})$ centred on $\boldsymbol{c}$. Because the annulus moves with the trial centre, the point reflection of every pixel in it also lies in the annulus. The annulus radii are fixed in physical units from the broad-band reference (Table~\ref{tab:setup}).

We locate the minimum on a 0.05-pixel grid, using cubic interpolation of the image, and refine it with a parabolic fit. The grid spacing sets only the numerical sampling and is not an uncertainty on the centre. The search window is centred on the broad-band lens position with a half-width of 1.5 PAUS pixels. If the minimum falls on the window edge, the half-width is increased to 3 and then 6 pixels. A minimum that still lies on the edge is rejected. Offsets of up to 0.15~arcsec are converted through the WCS and applied equally to the source and reflected-background regions. If the offset is larger, or no valid centre is found, the regions remain at the broad-band position, and the band is still measured rather than discarded.

All fluxes are measured on the observed images. Assuming the point-symmetric nature of early-type galaxies acting as lenses, the flux from the reflected background region is subtracted from the source flux. Let $w_{R,i}$ and $w_{P R,i}$
be the fractional pixel weights of the source and reflected-background
regions. With a local sky estimate $\bar{s}$, the source flux is
\begin{equation}
 \Farc=\sum_i(w_{R,i}-w_{P R,i})I_i
       -(A_R-A_{P R})\bar{s},
 \label{eq:estimator-full}
\end{equation}
where $A_R=\sum_i w_{R,i}$ and $A_{P R}=\sum_i w_{P R,i}$ are the effective areas of the two regions.
The sky term corrects any residual area mismatch. The lens flux is measured
in its own sky-subtracted region without reflected subtraction.
We sample each pixel with $81\times81$ subpixels and repeat the source
measurement with $161\times161$ sampling as a convergence check.

The analytic source variance includes the noise in both weighted regions,
a shared sky-level term $(A_R-A_{P R})^2{\rm Var}(\bar{s})$, and the Poisson
contribution when exposure information is available. Zero-point
uncertainties are propagated during calibration. Differential-noise measurements on surrounding blank regions are saved as diagnostics, while they do not replace the analytic errors in the reported fits. Valid exposures are combined by
inverse-variance weighting. Broad-band fluxes are rescaled from their survey
zero points to the PAUS convention of 26.

The cancellation is exact only for point-symmetric foreground light in the
lensed source region contaminated by the lens light and a reflected region containing pure lens light. PSF asymmetry,
unmasked companions, and source light in the reflected region can leave
residual biases. The estimator measures the source flux in the selected
image-plane region, not the total or demagnified source flux.

We then fit the lens and source photometry separately with BCNz \citep{eriksen2019}. Both branches use the same available bands: PAUS narrow bands together with HSC $grizy$ or CFHT $ugriz$. The retained set requires
finite fluxes and positive errors for both components.
We use the \citet{eriksen2019} template configuration, unit narrow-band
weight, and the stored grid $0.01\leq z\leq2.0$ with $\Delta z=0.001$.
BCNz adds its default 0.03-mag error floor and applies its default
$\mathrm{S/N}>-2$ validity cut. No additional spectroscopic zero-point
calibration is applied to these eight systems. Because the narrow- and broad-band fluxes are measured with different instruments and calibrations, BCNz fits their relative normalisation \citep{eriksen2019}. For each template run and trial redshift, the model broad-band fluxes are $\sum_s v_s\,t_s^{\rm BB}(z)$ and the model narrow-band fluxes are $k(z)\sum_s v_s\,t_s^{\rm NB}(z)$, where $t_s$ are the template fluxes and $v_s\geq0$ their coefficients. The coefficients are obtained by iterative non-negative minimisation of $\chi^2$ over all retained bands (1000 iterations). Every ten iterations, $k$ is reset to the ratio of the inverse-variance-weighted sums of the observed and model narrow-band fluxes, restricted to $0.1\leq k\leq10$. The scale is therefore fitted separately for each component, template run, and redshift, and the $\chi^2$ entering $p(z)$ is evaluated at its best-fitting value. The broad-band-only BCNz control (Sect.~\ref{sec:bb-only}) contains no narrow bands and hence no free scale.

\section{Results}
\label{sec:results}

\begin{figure*}[!tp]
\centering
\includegraphics[width=\textwidth]{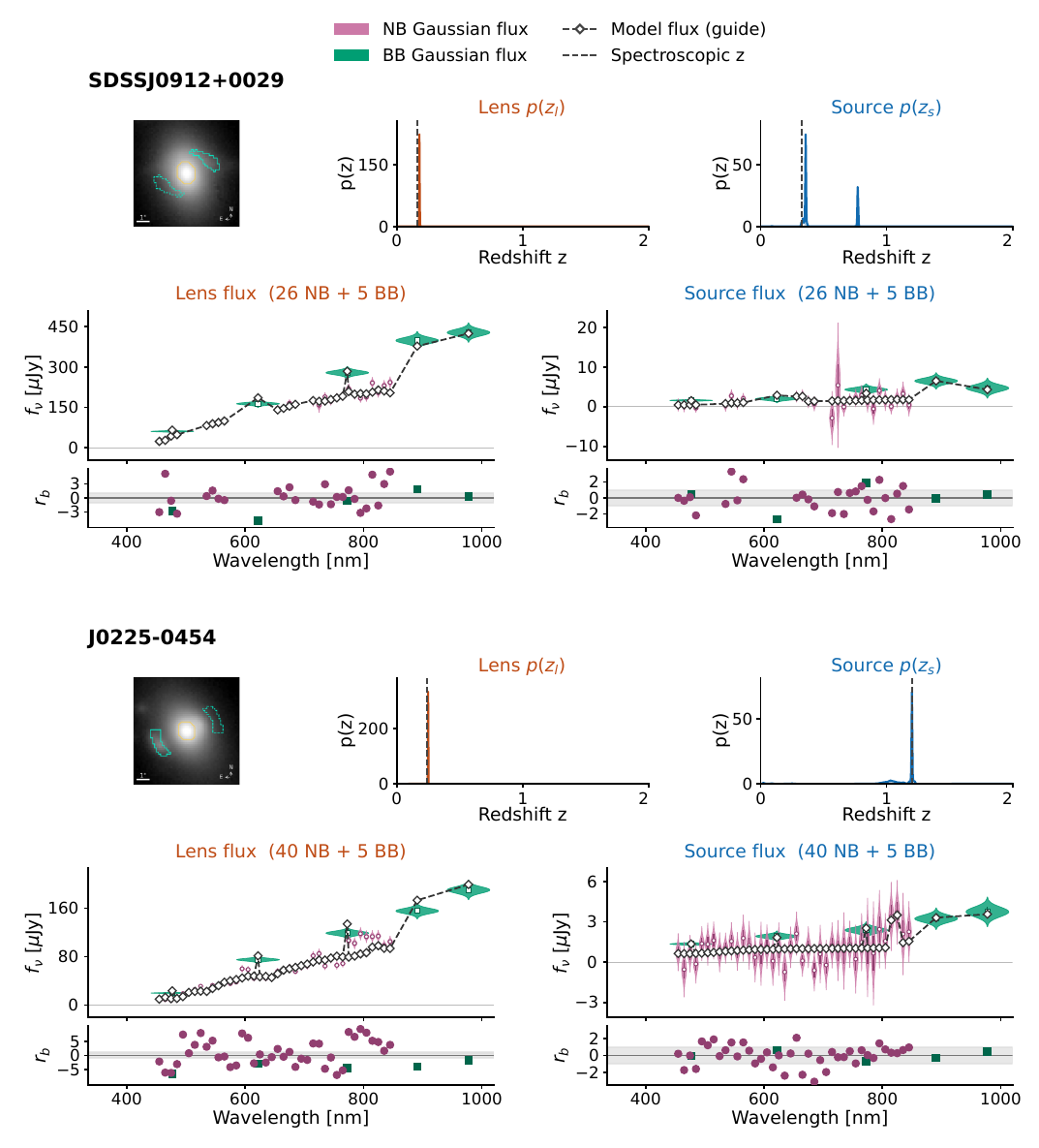}
\caption{Results for SDSSJ0912$+$0029 and J0225$-$0454. Each reference-image stamp spans $9\times9$~arcsec and shows source (solid cyan), lens (yellow), and reflected-background (dashed cyan) regions. Each system has three rows: the region image and lens/source $p(z)$ above, lens/source fluxes in the middle, and their aligned standardised residuals $r_b=(F_b-F_b^{\rm model})/\sigma_b$ below. Rose and green Gaussian symbols show narrow- and broad-band fluxes, extending to $\pm3\sigma$ with central $\pm1\sigma$ bars. Their fixed horizontal widths distinguish band classes. Open diamonds show model fluxes at the marginalised redshift peak, including the fitted NB/BB scale. Dashed connections in wavelength order are visual guides between band-integrated model values. The residual band marks $\pm1\sigma$, while shaded posterior regions indicate the central 68\% interval. The same layout and units will apply to Fig.~\ref{fig:results-b} and Appendix~\ref{app:other-systems}.}
\label{fig:results-a}
\end{figure*}

\begin{figure*}[!tp]
\centering
\includegraphics[width=\textwidth]{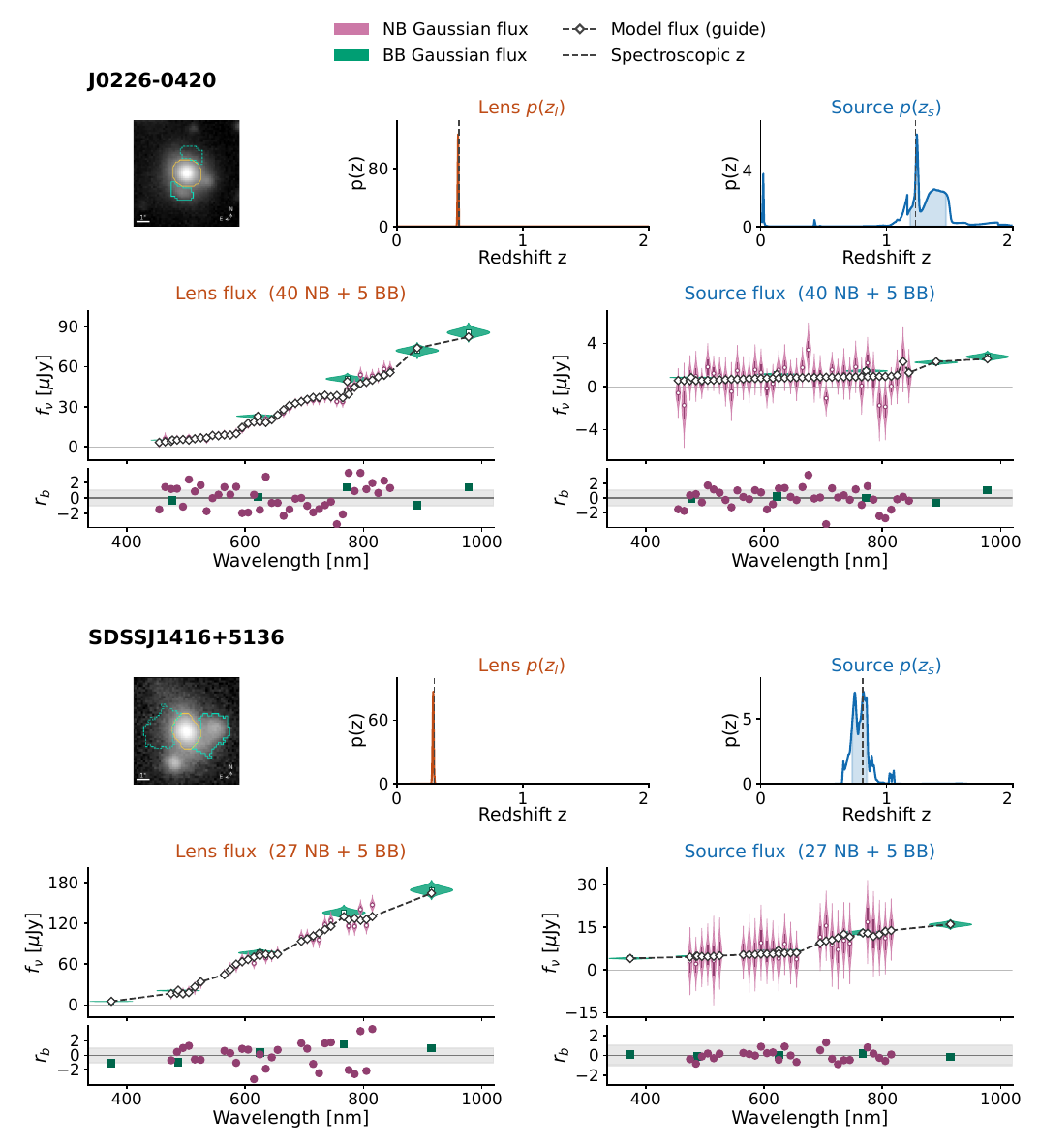}
\caption{Results for J0226$-$0420 and SDSSJ1416$+$5136, using the same three-row layout, Gaussian flux symbols,s dashed model connections, and residual display of Fig.~\ref{fig:results-a}.}
\label{fig:results-b}
\end{figure*}

In this section, we present the photometric-redshift results for the eight strong-lens systems. Section~\ref{sec:photo-z} compares the lens and source redshifts with spectroscopy, evaluates the redshift-ordering probability, and examines the posterior widths and fit quality. Section~\ref{sec:background-comparison} assesses how the symmetric-background subtraction affects the source redshifts, and Sect.~\ref{sec:bb-only} quantifies the information added by the narrow bands using broad-band-only controls. Finally, Sect.~\ref{sec:satellite-test} applies the same photometry to a neighbouring object to test whether it can be distinguished from the lensed source.

\subsection{Photometric redshifts and ordering}
\label{sec:photo-z}

We quantify the agreement between photometric and spectroscopic redshifts using the normalised residual $\delta_z=(z_b-z_{\rm spec})/(1+z_{\rm spec})$, where $z_b$ is the peak of the marginalised $p(z)$. The lens and source residuals are denoted by $\delta_{\rm l}$ and $\delta_{\rm s}$, respectively. We adopt $|\delta_z|<0.05$ as a common descriptive threshold for comparing all components and control tests. As shown in Table~\ref{tab:photoz}, all eight lens galaxies have $|\delta_{\rm l}|=0.002$--$0.014$. Four systems (J0225$-$0454, SDSSJ0912$+$0029, J0226$-$0420, and SDSSJ1416$+$5136) also show good agreement for their source redshifts, with $|\delta_{\rm s}|\leq0.025$. These systems are presented in Figures~\ref{fig:results-a} and \ref{fig:results-b}. The source redshift peaks of J0225$-$0454, J0226$-$0420, and SDSSJ1416$+$5136 are 1.201, 1.240, and 0.816, respectively, compared with spectroscopic redshifts of 1.20, 1.23, and 0.8115. Their corresponding normalised residuals are 0.0005, 0.0045, and 0.0025. SDSSJ0912$+$0029 shows a larger source residual of $\delta_{\rm s}=0.0250$. The remaining four systems have substantially larger source residuals, with $|\delta_{\rm s}|=0.117$--$0.636$, and are presented in Appendix~\ref{app:other-systems}. 
Figure~\ref{fig:all-eight} provides a comprehensive overview of the results for all eight systems.

In order to examine the ability of our method to provide effective redshift ordering information for vetting, we define the redshift-ordering probability as 
\begin{equation}
 P_{\rm ord}\equiv P(\zs>\zl)
 =\int p_{\rm l}(z_{\rm l})
       \left[\int_{z_{\rm l}}^{2.0}p_{\rm s}(z_{\rm s})\,{\rm d}z_{\rm s}\right]
       {\rm d}z_{\rm l}.
 \label{eq:ordering}
\end{equation}
All eight systems have $P_{\rm ord}>0.5$, indicating that the arcs surrounding the lens galaxies are more likely to originate from background sources than foreground objects. Five systems (J0225$-$0454, J0223$-$0534, J0226$-$0420, SDSSJ0912$+$0029, and SDSSJ1416$+$5136) have $P_{\rm ord}>0.9$ in Table~\ref{tab:photoz}. Their full redshift distributions strongly favour a source behind the lens under the adopted fitting model. In J0223$-$0534, the source peak at 1.158 misses the spectroscopic
value of 1.444, but $P_{\rm ord}=0.949$ still favours a source behind the lens. The redshift-ordering test can therefore provide useful information even when the source redshift is not accurately recovered.  

Although all eight lens peaks satisfy $|\delta_{\rm l}|\leq0.014$, seven of the lens spectroscopic redshifts lie outside the central 68\% credible interval of the BCNz posterior (quality code ``O'' in Table~\ref{tab:photoz}). The same applies to the source of SDSSJ0912$+$0029. The lens redshift posteriors are notably narrow, with their 68\% credible intervals spanning only 0.002--0.007 in units of $1+z$. The spectroscopic redshifts lie 0.002--0.012 beyond these intervals, indicating that the posterior uncertainties may be underestimated despite the agreement of their peaks with the spectroscopic values under our adopted criterion. Nevertheless, these offsets remain smaller than the effective wavelength resolution of the narrow-band photometry. A filter with FWHM $\simeq12.5$~nm at 455--845~nm corresponds to $\Delta\lambda/\lambda=\Delta z/(1+z)\simeq0.015$--0.027, and adjacent bands are separated by 0.012--0.022 in the same units. A continuum break or emission line can therefore be located only to within a fraction of a band width, and residuals of $\sim0.01$ are consistent with this sampling. The narrow posteriors arise because the bright lenses have high-S/N photometry, which makes the $\chi^2$ surface steep, whereas neither the 0.03-mag error floor nor the discrete template set fully accounts for template mismatch at this resolution. The large $\chi^2_{\min}/N$ values of several lenses (Table~\ref{tab:photoz}) support this interpretation. We therefore use the peak residual $\delta_z$ as the primary measure of agreement and regard the 68\% intervals as model-conditional rather than calibrated uncertainties. Calibrating them, for example with an additional template-error term, requires a larger spectroscopic sample.

The source $\chi^2/N_{\rm band}$ values, which quantify the agreement between the modelled and observed SEDs, range from 0.24 to 2.14. In contrast, J0225$-$0454 has a lens $\chi^2/N_{\rm band}$ of 22.10 despite its highly concentrated redshift posterior. This highlights that a narrow posterior does not necessarily imply a good fit to the observed SED. Template uncertainty can also produce very narrow photo-z
distributions \citep{wittman2016}. Likewise, $P_{\rm ord}$ is a
redshift-consistency diagnostic,
not a calibrated probability that a system is a strong lens.
Unlike joint fits of blends \citep{jones2019}, it neglects cross-component
covariance.

We find that increasing the region over-sampling from $81\times81$ to $161\times161$
subpixels changes every tested source flux by less than $0.122$ times its
adopted measurement error. This establishes numerical convergence of the
integration. It does not establish that the foreground subtraction is
unbiased or account for uncertainty in the region geometry.

\begin{table*}[t]
\caption{Eight-system redshift summary. L and S denote lens and source. NB/BB counts are the final bands used by BCNz. Quality codes combine peak agreement (A: $|\delta_z|<0.05$; D: larger discrepancy) and interval inclusion (I/O: spectroscopy inside/outside the central 68\% credible interval). The intervals are supplied with the data products and are conditional on the model and $0.01\leq z\leq2$ grid. BCNz summarises these two diagnostics for the present sample. $\chi^2_{\min}/N$ is the global minimum per retained band, not reduced $\chi^2$. $P_{\rm ord}$ is defined in Eq.~(\ref{eq:ordering}).}
\label{tab:photoz}
\centering\footnotesize
\setlength{\tabcolsep}{2.3pt}
\begin{tabular}{llrrrrrrrr}
\toprule
System & Part & NB/BB & $z_{\rm spec}$ & $z_b$ & Central 68\% & $\delta_z$ & $\chi^2_{\min}/N$ & Quality & $P_{\rm ord}$ \\
\midrule
SDSSJ0912$+$0029 & L & 26/5 & 0.1642 & 0.180 & [0.178, 0.182] & $+0.0136$ & 6.19 & A/O & 0.997 \\
 & S & 26/5 & 0.3239 & 0.357 & [0.352, 0.770] & $+0.0250$ & 2.14 & A/O &  \\
\addlinespace[2pt]
J0225$-$0454 & L & 40/5 & 0.2379 & 0.249 & [0.248, 0.250] & $+0.0089$ & 22.10 & A/O & 0.970 \\
 & S & 40/5 & 1.2000 & 1.201 & [1.027, 1.204] & $+0.0005$ & 1.32 & A/I &  \\
\addlinespace[2pt]
J1404$+$5200 & L & 40/5 & 0.4560 & 0.449 & [0.447, 0.451] & $-0.0048$ & 4.10 & A/O & 0.513 \\
 & S & 40/5 & 1.5900 & 0.241 & [0.198, 1.808] & $-0.5208$ & 1.05 & D/I &  \\
\addlinespace[2pt]
SL2SJ0217$-$0513 & L & 39/5 & 0.6462 & 0.640 & [0.631, 0.642] & $-0.0037$ & 3.66 & A/O & 0.653 \\
 & S & 39/5 & 1.8500 & 0.037 & [0.059, 1.858] & $-0.6361$ & 0.93 & D/I &  \\
\addlinespace[2pt]
J0223$-$0534 & L & 40/5 & 0.4990 & 0.493 & [0.489, 0.496] & $-0.0040$ & 2.88 & A/O & 0.949 \\
 & S & 40/5 & 1.4440 & 1.158 & [1.030, 1.645] & $-0.1170$ & 1.24 & D/I &  \\
\addlinespace[2pt]
J0226$-$0420 & L & 40/5 & 0.4940 & 0.487 & [0.483, 0.490] & $-0.0047$ & 2.47 & A/O & 0.964 \\
 & S & 40/5 & 1.2300 & 1.240 & [1.186, 1.472] & $+0.0045$ & 1.57 & A/I &  \\
\addlinespace[2pt]
J1406$+$5226 & L & 40/5 & 0.7160 & 0.712 & [0.708, 0.718] & $-0.0023$ & 2.93 & A/I & 0.585 \\
 & S & 40/5 & 1.4700 & 0.480 & [0.428, 1.853] & $-0.4008$ & 1.15 & D/I &  \\
\addlinespace[2pt]
SDSSJ1416$+$5136 & L & 27/5 & 0.2987 & 0.289 & [0.284, 0.293] & $-0.0075$ & 2.69 & A/O & $>0.999$ \\
 & S & 27/5 & 0.8115 & 0.816 & [0.726, 0.844] & $+0.0025$ & 0.24 & A/I &  \\
\bottomrule
\end{tabular}

\end{table*}

\begin{figure*}[t]
\centering
\includegraphics[width=\dimexpr\textwidth-2\fboxsep-2\fboxrule\relax]{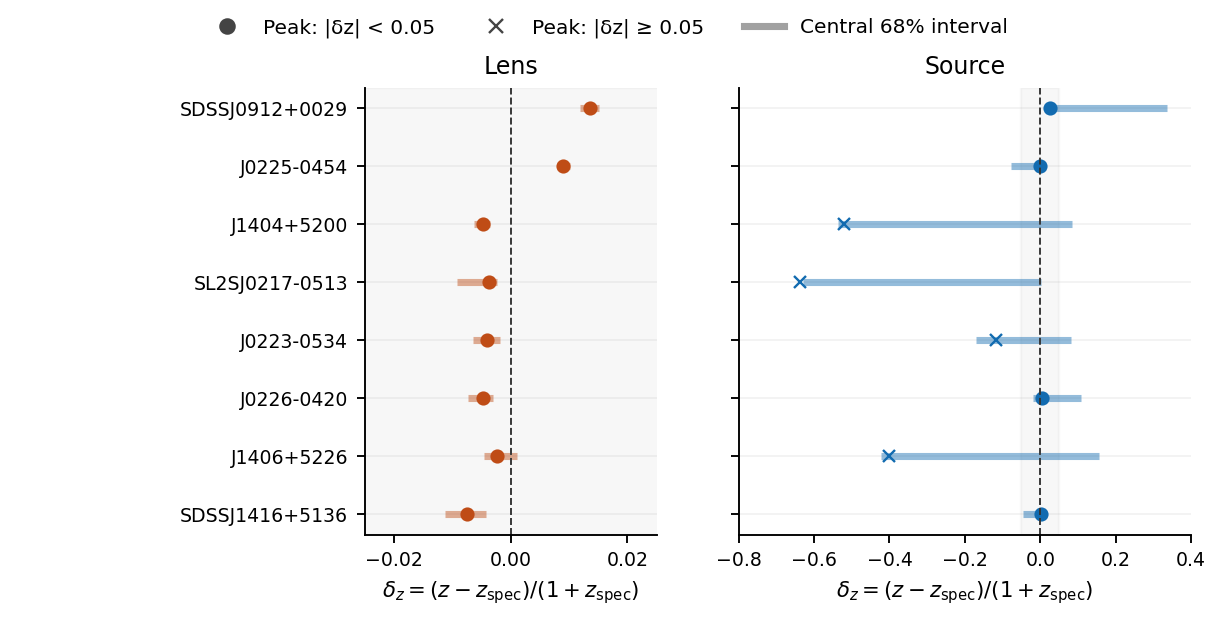}
\caption{All eight known lenses, including the four large source-peak discrepancies. Points mark redshift peaks, while horizontal segments give the central 68\% interval expressed relative to the spectroscopic value. Segments need not be centred on the peak. Circles and crosses distinguish $|\delta_z|<0.05$ and $\geq0.05$, respectively. The two panels use different horizontal scales to show the small lens offsets and large source uncertainties.}
\label{fig:all-eight}
\end{figure*}

\subsection{Comparison without symmetric-background subtraction}
\label{sec:background-comparison}

We repeat the source extraction without subtracting the point-reflected
region. The source region, accepted exposures, and centre corrections
are held fixed. The control still subtracts the local sky background, but retains
foreground light within the source region without symmetric subtraction. Its noise and exposure weights are recalculated for this measurement. Both fits use the same templates,
error floor, and quality cuts. The actual retained bands are also identical
for all eight paired comparisons. The lens photometry is unchanged.

Our method mainly affects the measurement of the source redshift. Table~\ref{tab:background-comparison} shows that symmetric subtraction reduces the absolute source-redshift bias in five systems. The median $|\delta_{\rm s}|$ falls from 0.336 to 0.071. Without symmetric subtraction, no source is recovered within $|\delta_{\rm s}|<0.05$, and all eight source peaks lie below their spectroscopic
redshifts. In J0223$-$0534 and J1406$+$5226, the peaks of 0.493 and 0.715 nearly
coincide with the lens peaks of 0.493 and 0.712. This behaviour is consistent with foreground contamination dominating the inferred colours. Such colour biases also affect model-based source photometry \citep{langeroodi2023}.

For J0226$-$0420, subtraction moves the peak from 0.619 to 1.240, close to the spectroscopic value of 1.23. The unsubtracted fit even gives $P_{\rm ord}=0.997$ despite its incorrect redshift. Symmetric subtraction can thus improve the accuracy of the source redshift, although residual foreground contamination can still affect it.

The comparison shows that changing only the foreground treatment, with all other inputs fixed, can significantly affect the source redshift estimate. The symmetric-background subtraction is therefore a crucial step in our pipeline to mitigate foreground contamination and improve the accuracy of the source photometric redshift.

\subsection{Broad-band-only control}
\label{sec:bb-only}

To quantify the information added by the narrow bands, we fit the same
aperture fluxes using only the broad bands. Broad-band photometric redshifts
are usually estimated either by template fitting, for example with BPZ
\citep{benitez2000}, LePhare \citep{arnouts1999,ilbert2006}, EAZY
\citep{brammer2008}, or by machine-learning methods trained on spectroscopic
samples \citep[e.g.][]{collister2004,sadeh2016}. Both approaches are used for HSC-SSP \citep{tanaka2018}, and template
fitting has been used for CFHTLS \citep{ilbert2006,hildebrandt2012}.
Machine-learning estimators require training samples that resemble the targets,
which is not the case for lens-subtracted aperture fluxes of arcs. We
therefore use the template-fitting code EAZY, with its default set of FSPS templates \citep{conroy2009}. The inputs are the production HSC $grizy$ orCFHT $ugriz$ fluxes used in the PAUS NB+BB fits, including the symmetric-background subtraction for the source. We add a 0.03-mag systematic error, as in BCNz, and use the same grid, $0.01\leq z\leq2.0$ with $\Delta z=0.001$. We switch off the apparent-magnitude prior, because the arcs are magnified and the regions contain only a part of each galaxy. As a second control, we repeat the BCNz fit with the same templates but without the narrow bands.

\begin{table*}[t]
\caption{Redshift peaks from the production PAUS NB+BB fit, the PAUS NB+BB control without symmetric-background subtraction, and the two broad-band-only controls, all using the same region geometry. The ``No subtraction'' column gives the source photometric-redshift peak inferred by BCNz from PAUS NB+BB fluxes measured in the source region without point-reflected foreground-light subtraction, while the local sky background is still removed. Daggers mark $|\delta_z|\geq0.05$ and the final row counts $|\delta_z|<0.05$, using the same normalised residual as Table~\ref{tab:photoz}.}
\label{tab:bb-only}
\label{tab:background-comparison}
\centering
\footnotesize
\setlength{\tabcolsep}{3pt}
\begin{tabular}{lrrrrrrrrr}
\toprule
 & \multicolumn{4}{c}{Lens} & \multicolumn{5}{c}{Source} \\
\cmidrule(lr){2-5}\cmidrule(lr){6-10}
System & $z_{\rm spec}$ & PAUS+BB & EAZY BB & BCNz BB & $z_{\rm spec}$ & PAUS+BB & No subtraction & EAZY BB & BCNz BB \\
\midrule
SDSSJ0912$+$0029 & 0.1642 & 0.180 & 0.300$^{\dagger}$ & 0.746$^{\dagger}$ & 0.3239 & 0.357 & 0.170$^{\dagger}$ & 0.838$^{\dagger}$ & 1.267$^{\dagger}$ \\
J0225$-$0454 & 0.2379 & 0.249 & 0.451$^{\dagger}$ & 0.391$^{\dagger}$ & 1.2000 & 1.201 & 0.306$^{\dagger}$ & 0.298$^{\dagger}$ & 0.422$^{\dagger}$ \\
J1404$+$5200 & 0.4560 & 0.449 & 0.355$^{\dagger}$ & 0.612$^{\dagger}$ & 1.5900 & 0.241$^{\dagger}$ & 0.642$^{\dagger}$ & 0.021$^{\dagger}$ & 0.430$^{\dagger}$ \\
SL2SJ0217$-$0513 & 0.6462 & 0.640 & 0.746$^{\dagger}$ & 0.671 & 1.8500 & 0.037$^{\dagger}$ & 0.787$^{\dagger}$ & 0.061$^{\dagger}$ & 0.047$^{\dagger}$ \\
J0223$-$0534 & 0.4990 & 0.493 & 0.477 & 0.691$^{\dagger}$ & 1.4440 & 1.158$^{\dagger}$ & 0.493$^{\dagger}$ & 1.495 & 0.434$^{\dagger}$ \\
J0226$-$0420 & 0.4940 & 0.487 & 0.477 & 0.594$^{\dagger}$ & 1.2300 & 1.240 & 0.619$^{\dagger}$ & 1.994$^{\dagger}$ & 0.423$^{\dagger}$ \\
J1406$+$5226 & 0.7160 & 0.712 & 0.627$^{\dagger}$ & 0.651 & 1.4700 & 0.480$^{\dagger}$ & 0.715$^{\dagger}$ & 0.383$^{\dagger}$ & 0.822$^{\dagger}$ \\
SDSSJ1416$+$5136 & 0.2987 & 0.289 & 0.352 & 0.514$^{\dagger}$ & 0.8115 & 0.816 & 0.456$^{\dagger}$ & 0.778 & 0.762 \\
\midrule
$|\delta_z|<0.05$ &  & 8/8 & 3/8 & 2/8 &  & 4/8 & 0/8 & 2/8 & 1/8 \\
\bottomrule
\end{tabular}

\end{table*}

Table~\ref{tab:bb-only} shows that removing the narrow bands degrades both components. With the narrow bands, eight lenses and four sources are recovered within $|\delta_z|<0.05$. The broad-band-only EAZY fit recovers three lens and two source redshifts, and the broad-band-only BCNz fit recovers two and one. The median $|\delta_{\rm l}|$ increases from 0.005 with the narrow bands to 0.056 for EAZY and 0.115 for BCNz. Three of the four sources recovered with the narrow bands are not recovered by either broad-band-only fit (SDSSJ0912$+$0029, J0225$-$0454, and J0226$-$0420). Only SDSSJ1416$+$5136 is recovered in all three fits. In the opposite direction, EAZY recovers the source of J0223$-$0534 at 1.495, whereas the PAUS NB+BB fit gives 1.158, suggesting that the BCNz template configuration could be improved for such sources.

\subsection{A neighbouring-object redshift diagnostic}
\label{sec:satellite-test}

\begin{figure}[h]
\centering
\includegraphics[width=\columnwidth]{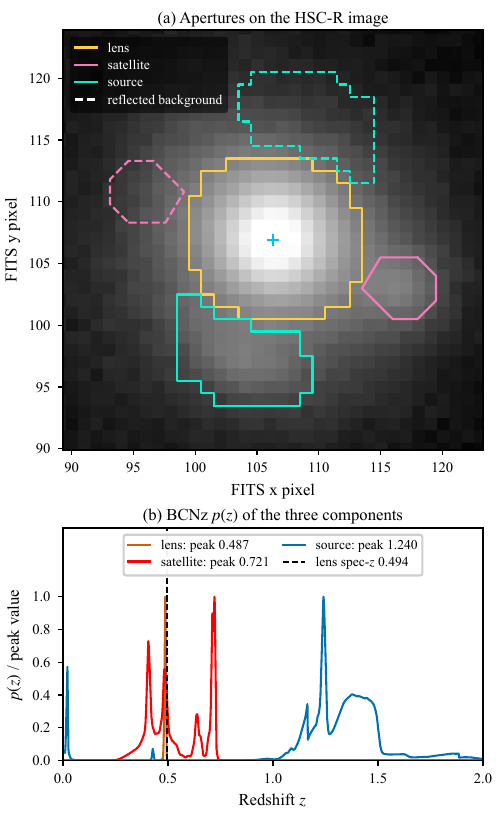}
\caption{Photometric redshift comparison for the source and a neighbouring object in J0226$-$0420. The upper panel shows the lens-subtracted image with the source (cyan) and satellite (magenta) regions. The lower panel shows the normalised redshift distributions of the two objects. The source posterior peaks at 1.240, while the satellite posterior peaks at 0.721 and has a broad distribution that includes the lens redshift of 0.494. Treating the two posteriors as independent, $P(z_{\rm sat}<z_{\rm s})=0.965$ and their overlap coefficient is 0.005. Under the adopted model, the object is therefore unlikely to share the arc's redshift, but its broad posterior does not establish physical association with the lens.}
\label{fig:satellite-test}
\end{figure}

We extend the component-photometry analysis to a neighbouring object to explore the redshift information available for structures around the lens.

J0226$-$0420 provides a test of component classification beyond the main arc.
We extract the object labelled as a ``satellite'' light component in Fig.~\ref{fig:satellite-test}
using its own region and a counterpart reflected about the lens-light centre. The reflected region does not geometrically overlap the lens, source, or satellite regions.  The neighbouring object fit uses 38 narrow bands (NB625 and NB645 are excluded as noisy) and five broad bands, without assigning the source spectroscopic redshift to this object.

The neighbouring object redshift posterior has a peak at 0.721 and a median of 0.492.
Its central 68\,per cent interval is $[0.404,0.712]$, which includes the
lens spectroscopic redshift of 0.494. The ordering probability is
$P(z_{\rm sat}>z_{\rm l})=0.537$.
By contrast, the source redshift posterior has a peak at 1.240, a central
68\,per cent interval of $[1.186,1.472]$, and $P_{\rm ord}=0.964$.
Assuming independent posteriors, the probability that the neighbouring object lies in front of the source is $P(z_{\rm sat}<z_{\rm s})=0.965$, and the overlap coefficient $\int\min[p_{\rm sat}(z),p_{\rm s}(z)]\,{\rm d}z$ is 0.005. This assumption neglects photometric systematics shared by the two regions.
The neighbouring object's distribution favours lower redshifts than the arc's, illustrating the additional information available from separate component photometry. Its broad posterior includes the lens redshift, making physical association with the lens a hypothesis for spectroscopic follow-up. Quantifying the relative support for a lens association or a common redshift with the arc requires the full posterior and a joint treatment of shared photometric systematics.

\section{Discussion}
\label{sec:discussion}

\subsection{Conditions for recovering both redshifts}

SDSSJ0912$+$0029, SDSSJ1416$+$5136, J0225$-$0454, and J0226$-$0420 recover both redshift peaks reasonably well. The lens redshifts place the 4000~\AA\ break at 466--686~nm across the full sample, within the PAUS wavelength range. This feature provides a natural redshift constraint for early-type galaxies \citep{eriksen2019,renard2022d4000}. The sources are fainter and have more diverse spectral shapes, so their redshifts depend more strongly on the available features and band coverage \citep{salvato2019}.

J0225$-$0454 provides the clearest indication of an emission-line contribution of the source.
At $z_{\rm s}^{\rm spec}=1.20$, [O\,II]~$\lambda3727$ falls at 819.9~nm.
The excess in NB815 and NB825 is consistent with this location and with the
best-fitting model. The measured flux S/N values are 4.0 and 4.7, respectively.  These are not continuum-subtracted line significances.
For J0226$-$0420, [O\,II] is expected at 831.1~nm, but NB835 has a flux S/N of only 2.5. Its accurate peak therefore cannot be attributed to a
secure line detection. Broad-band colours and the combined weak narrow-band
constraints may instead determine the preferred solution.

The distinction is stronger for SDSSJ0912$+$0029 and
SDSSJ1416$+$5136. Their expected
[O\,II] wavelengths are 493.4 and 675.1~nm, but the nearest narrow bands
are missing from the retained photometry. Their recovery cannot be
attributed to a well-sampled [O\,II] feature. Continuum information, including
the 4000~\AA\ break at 529.6 and 724.6~nm, offers alternative constraints.
All four better-recovered sources have $z_{\rm s}^{\rm spec}\leq1.23$.
The other four have $z_{\rm s}^{\rm spec}\geq1.41$, placing both [O\,II]
and the 4000~\AA\ break beyond the nominal PAUS narrow-band range.
This pattern suggests that spectral coverage matters, but does not establish
a universal redshift limit. Broad bands can still constrain higher-redshift
sources, as the broad-band-only EAZY fit to J0223$-$0534 shows
(Sect.~\ref{sec:bb-only}). The same comparison shows that the narrow bands are
needed for the other three recovered sources. Fits that omit selected narrow
bands are still needed to identify which spectral features drive each result.

Foreground removal is also important. None of the four source peaks is
recovered within $|\delta_{\rm s}|<0.05$ in the unsubtracted control.
Successful measurements require useful narrow-band photometry information and sufficiently
clean differential photometry, not merely a large number of bands.
Residual asymmetry, PSF wings, and errors in the adopted centre can bias the
colours. The present noise model does not propagate uncertainty in that
centre or in the region boundaries.  

\subsection{Integration into lens-candidate validation}

Strong-lens validation connects candidate discovery, expert grading,
spectroscopic confirmation, and subsequent scientific analysis. Current
searches combine automated selection with visual inspection and targeted
follow-up \citep{wong2022,jaelani2024,gonzalez2026}. The purpose of our method
is to supply additional information throughout this process, rather than to
replace spectroscopic redshifts. Its main outputs, the separate lens and
source distributions $p_{\rm l}(z)$ and $p_{\rm s}(z)$ and their ordering
probability $P_{\rm ord}$, can support different decisions at each stage.

After an imaging search has identified a list of candidates, component-level narrow-band photometry can help prioritise them for further inspection. A source redshift distribution favouring solutions behind the deflector provides additional support for promising candidates, whereas a low-redshift solution may indicate a satellite galaxy or foreground interloper. During expert grading, our method can complement morphological and colour information by providing photometric redshift distributions and $P_{\rm ord}$. As illustrated by the neighbouring-object test in Sect.~\ref{sec:satellite-test}, this component-level information can help distinguish potential lensed sources from satellite galaxies or interlopers.

The same information can guide spectroscopic follow-up. The source
redshift distribution predicts a range of observed wavelengths for candidate
emission lines through $\lambda_{\rm obs}=(1+z_{\rm s})\lambda_{\rm rest}$.
Together with the arc brightness, this can inform target priorities,
instrument selection, wavelength coverage, and exposure planning. A multimodal photo-$z$ distribution would motivate
coverage of several possible line locations. Spectroscopy remains the reference for precise component redshifts and for calibrating the photometric estimates, while the photo-$z$ contribution is to make spectroscopy targeting and interpretation better informed.

Beyond validation, the redshift information can also enter scientific analyses of high-probability systems that lack spectroscopic follow-up. For an assumed cosmology, a joint lens--source redshift distribution can be propagated into a distribution of the angular-diameter distance ratio $D_{\rm ls}/D_{\rm s}$, where $D_{\rm ls}$ and $D_{\rm s}$ are the lens-to-source and observer-to-source distances. Retaining the full distribution, this distance-ratio distribution could then be used to constrain lens masses or to plan cosmological follow-up at the population level. \citet{sonnenfeld2022} demonstrated that galaxy-structure inference can marginalise over unknown source redshifts when the source population and lens selection function are modelled. Our component photometry could provide additional constraints for studies of mass profiles and galaxy evolution in lens samples with incomplete spectroscopic coverage.

Realising these uses requires tests on candidate lists containing both
lenses and non-lenses, together with simulated blended systems, to measure
false-positive rates and selection effects \citep{metcalf2019}. The present
sample consists of known lenses selected using spectroscopic redshifts. These redshifts enter neither the broad-band models, which adopt fiducial values (Sect.~\ref{sec:model}), nor the photo-$z$ fits. A blind
application must therefore test region selection without known source
redshifts, calibrate the redshift distributions against spectroscopy, and
account for shared photometric uncertainties and the current $z=2$ grid
boundary. These tests would establish how much information the method adds
at each stage, from candidate ranking to statistical science.

\subsection{Future prospects}

Future applications can benefit from larger lens samples, broader
wavelength coverage, and improved spatial resolution. Euclid's first lens
catalogue demonstrates its capacity to discover large samples and resolve
systems with small Einstein radii \citep{walmsley2026}. Its imaging could
help define component regions for narrow- or medium-band measurements in
overlapping fields. Such combinations would extend our approach, provided
that the ground-based data have sufficient depth and that PSF differences
are accounted for.

J-PAS offers a direct extension of the PAUS approach through its
planned coverage of at least $8000~\mathrm{deg}^{2}$ and 54 narrow bands
spanning approximately 3780--9100~\AA\
\citep{benitez2014,bonoli2021}\footnote{J-PAS survey overview:
\url{https://www.j-pas.org/survey}.}. Its larger area could increase the
available lens sample, while its redder coverage would extend the nominal
upper redshift for [O\,II] to about 1.44 and for the 4000~\AA\ break to
about 1.28. Reliable constraints at these limits would still require
adequate photometric S/N.

Deep broad- and medium-band data provide complementary constraints.
Combining miniJPAS and HSC-SSP redshift distributions has already improved
photo-$z$ estimates \citep{hernanCaballero2024}. The 16 HSC-Niji medium
bands, with central wavelengths of approximately 413--960~nm, could
strengthen continuum-break constraints beyond the PAUS red limit
\citep{euclidTarsitano2026}\footnote{HSC-Niji survey and filter design:
\url{https://hsc-niji.github.io/}.}. Deep broad-band imaging from HSC and
Rubin's LSST could similarly be combined with medium- or narrow-band
observations in overlapping fields or targeted follow-up
\citep{bosch2018,ivezic2019}.

The Multi-Channel Imager (MCI) on the Chinese Space Station Survey
Telescope (CSST) offers simultaneous observations in three channels
spanning 255--1000~nm, with 30 broad-, medium-, and narrow-band filters and
a $0.05$-arcsec pixel scale \citep{mci2025}. Near-ultraviolet data could
help distinguish competing source redshift solutions, while the red
channel would sample features beyond the PAUS window. The red limit
corresponds to a nominal [O\,II] redshift of $z\simeq1.68$. MCI's mixed filter
set would complement the denser optical sampling of PAUS or J-PAS. 

Our work serves as a proof of concept for applying narrow-band photometry to photometric redshift estimation in strong-lensing systems. The improvements achievable with next-generation instruments still need to be quantitatively evaluated using their actual filter transmission curves and spectroscopic validation samples, while accounting for photometric signal-to-noise ratios, wavelength-dependent PSFs, and residuals from foreground-light subtraction.

\subsection{Beyond strong lensing}

The extraction does not require the faint object to be gravitationally
lensed. Its essential assumption is that the foreground light can be
estimated from a point-reflected region. The relatively low asymmetry
of early-type galaxies makes them suitable foregrounds
\citep{conselice2000,conselice2003}. The same principle can therefore be
applied to close projected galaxies, satellite candidates, and faint
companions embedded in the light of a bright early-type galaxy.

For such applications, image-based component modelling can replace the
lens-specific model used to define the regions here. Each neighbouring
object would have its own differential fluxes and redshift distribution,
as in the satellite test for J0226$-$0420. This could improve colour and
redshift estimates when ordinary aperture photometry is dominated by the
foreground, and help separate likely companions from background projections.

This application is complementary to established approaches to blended
photometry. T-PHOT uses high-resolution spatial priors to estimate fluxes
in lower-resolution images \citep{merlin2015}. Joint photometric-redshift
inference can also separate components statistically, including cases that
are only resolved in some bands \citep{jones2019}. Our approach instead
uses local foreground symmetry to extract aperture fluxes before fitting
the redshifts.

The limits of applying this method to the photometry of ordinary early-type galaxies should also be recognised. The object must be sufficiently resolved, the reflected region must be free of other sources, and the
observed foreground must be sufficiently symmetric. Dust, tidal structure,
asymmetric PSFs, or source light in the reflected region can violate
these assumptions. Adequate S/N and spectral coverage are also required.
The present lens sample motivates this broader use, but a dedicated
spectroscopic validation is needed before claiming high redshift accuracy
for the general population of close neighbours.

\section{Conclusions}
\label{sec:conclusions}

We have presented a proof-of-concept demonstration of component photometric redshifts in eight known strong lenses, combining broad-band lens modelling, symmetric-region foreground subtraction, and separate BCNz fits to NB+BB photometry. All eight lens peaks have $|\delta_{\rm l}|\leq0.014$ and four source peaks have $|\delta_{\rm s}|\leq0.025$. The four closer source matches occur at $z_{\rm s}^{\rm spec}\leq1.23$, while the other four source residuals are 0.117--0.636 in absolute value. These results demonstrate that useful redshift information can be recovered from blended lens and source light in favourable cases by symmetric-region foreground subtraction.

At the common descriptive threshold $|\delta_z|<0.05$, PAUS NB+BB matches 8/8 lens and 4/8 source redshifts, compared with 3/8 and 2/8 for EAZY BB-only, and 2/8 and 1/8 for BCNz BB-only. Together, these comparisons support the value of foreground subtraction and narrow-band information for component redshift estimation in this sample.

The present evidence establishes feasibility and quantifies best-fit agreement in known lenses. The posterior intervals still require independent calibration, and candidate-validation performance remains to be tested on samples containing both lenses and non-lenses. Larger spectroscopic samples, image simulations, and joint propagation of photometric and modelling uncertainties are the next steps towards assessing the reliability and wider applicability of the method.

Looking ahead, wider narrow-band surveys such as J-PAS, together with deep broad-band imaging and complementary medium-band observations, offer opportunities to apply and test this approach across larger and more diverse samples. With these validation steps, component redshift constraints could complement morphology and colour in prioritising spectroscopic follow-up and support studies of lens populations with incomplete spectroscopy. The present demonstration provides a practical starting point for developing this use of multi-band imaging in strong-lens research.

\begin{acknowledgements}
PR acknowledges support from the Spanish Ministerio de Ciencia, Innovación y Universidades, through projects PID2022-138896NB; and the programme Unidad de Excelencia María de Maeztu, financed by MCIN/AEI/10.13039/501100011033, and by the MaX-CSIC Excellence Award MaX4-SOMMA-ICE. 
QH was supported by the Dutch Research Council (NWO) through the grant Open Competitie ENW M21-1.
DNG acknowledges support from the European Research Council (ERC) under the European Union’s Horizon 2020 research and innovation programme with Grant agreement No. 101053992.
MG acknowledges support from the Polish National Agency for Academic Exchange grant No. BPN/BEK/2024/1/00325/DEC/1. 

This work uses imaging data from the Physics of the Accelerating Universe Survey (PAUS), obtained with PAUCam on the William Herschel Telescope. The PAU Survey has received support from the Spanish Ministry of Economy and Competitiveness (MINECO), including grants CSD2007-00060, AYA2015-71825, ESP2017-89838, PGC2018-094773, PGC2018-102021, PID2019-111317GB, SEV-2016-0588, SEV-2016-0597, and MDM-2015-0509, as well as European Union funding through the LACEGAL and EWC Marie Sklodowska-Curie programmes (grant agreements 734374 and 776247). Additional support has been provided by Durham University, ETH Zurich, Leiden University, University College London, and associated European Research Council and national funding programmes. The IEEC and IFAE receive institutional support through the CERCA programme of the Generalitat de Catalunya.

PAUS data processing and storage are supported by the Port d'Informacio Cientifica (PIC), operated through a collaboration between CIEMAT and IFAE, with additional contributions from the Universitat Autonoma de Barcelona and the European Regional Development Fund. We acknowledge the PIC team for providing the data infrastructure and technical support required for this work.

The Hyper Suprime-Cam (HSC) collaboration includes the astronomical communities of Japan and Taiwan, and Princeton University. The HSC instrumentation and software were developed by the National Astronomical Observatory of Japan (NAOJ), the Kavli Institute for the Physics and Mathematics of the Universe (Kavli IPMU), the University of Tokyo, the High Energy Accelerator Research Organization (KEK), the Academia Sinica Institute for Astronomy and Astrophysics in Taiwan (ASIAA), and Princeton University. Funding was contributed by the FIRST program from the Japanese Cabinet Office, the Ministry of Education, Culture, Sports, Science and Technology (MEXT), the Japan Society for the Promotion of Science (JSPS), Japan Science and Technology Agency (JST), the Toray Science Foundation, NAOJ, Kavli IPMU, KEK, ASIAA, and Princeton University. 

This paper makes use of software developed for Vera C. Rubin Observatory. We thank the Rubin Observatory for making their code available as free software at \url{http://pipelines.lsst.io/}.

This paper is based on data collected at the Subaru Telescope and retrieved from the HSC data archive system, which is operated by the Subaru Telescope and Astronomy Data Center (ADC) at NAOJ. Data analysis was in part carried out with the cooperation of Center for Computational Astrophysics (CfCA), NAOJ. We are honored and grateful for the opportunity of observing the Universe from Maunakea, which has the cultural, historical and natural significance in Hawaii. 
\end{acknowledgements}

\section*{Data availability}
\label{sec:data-avail}
The data products and reproduction scripts for all eight systems will be made publicly available on GitHub. The release will include machine-readable fluxes before and after the BCNz error-floor step, errors, common and component-specific band masks, exclusion reasons, fitted model fluxes, standardised residuals, normalised $p(z)$, and central 68\%/95\% intervals. It will also include reference-band/PSF metadata, PSF stamps, aperture polygons and areas, per-band centring records, accepted source-exposure diagnostics and calibration factors, control-fit peaks, and input SHA-256 hashes. The reproduction scripts will regenerate the tables and figures from these saved products without refitting or changing the models. 

\bibliographystyle{aa}
\bibliography{references}

@ARTICLE{he2024,
  author  = {He, Qiuhan and Nightingale, James W. and Amvrosiadis, Aristeidis
             and Robertson, Andrew and Cole, Shaun and Frenk, Carlos S. and
             Massey, Richard and Li, Ran and Cao, Xiaoyue and
             Lange, Samuel C. and Fran\c{c}a, Jo\~{a}o Paulo C.},
  title   = {Unveiling lens light complexity with a novel multi-Gaussian
             expansion approach for strong gravitational lensing},
  journal = {MNRAS},
  year    = {2024},
  volume  = {532},
  number  = {2},
  pages   = {2441},
  doi     = {10.1093/mnras/stae1577}
}

@article{desi2022,
    author  = {{DESI Collaboration} and others},
    title   = {Overview of the Instrumentation for the Dark Energy Spectroscopic Instrument},
    journal = {AJ},
    year    = {2022},
    volume  = {164},
    number  = {5},
    pages   = {207},
    doi     = {10.3847/1538-3881/ac882b},
    adsurl  = {https://ui.adsabs.harvard.edu/abs/2022AJ....164..207D}
}

@article{huang2025desi,
    author = {{Huang}, Xiaosheng and {Inchausti}, Jose Carlos and {Storfer}, Christopher J. and others},
    title = {{DESI Strong Lens Foundry II: DESI Spectroscopy for Strong Lens Candidates}},
    journal = {arXiv e-prints},
    year = {2025},
    eid = {arXiv:2509.18089},
    eprint = {2509.18089},
    archivePrefix = {arXiv},
    primaryClass = {astro-ph.CO},
    doi = {10.48550/arXiv.2509.18089}
}

@inproceedings{vaswani2017,
    author    = {Vaswani, Ashish and Shazeer, Noam and
                 Parmar, Niki and Uszkoreit, Jakob and
                 Jones, Llion and Gomez, Aidan N. and
                 Kaiser, Lukasz and Polosukhin, Illia},
    title     = {Attention Is All You Need},
    booktitle = {Advances in Neural Information Processing Systems},
    volume    = {30},
    pages     = {5998--6008},
    year      = {2017},
    eprint    = {1706.03762},
    archivePrefix = {arXiv}
}

@inproceedings{dosovitskiy2021,
    author    = {Dosovitskiy, Alexey and Beyer, Lucas and
                 Kolesnikov, Alexander and Weissenborn, Dirk and
                 Zhai, Xiaohua and Unterthiner, Thomas and
                 Dehghani, Mostafa and Minderer, Matthias and
                 Heigold, Georg and Gelly, Sylvain and
                 Uszkoreit, Jakob and Houlsby, Neil},
    title     = {An Image Is Worth 16x16 Words:
                 Transformers for Image Recognition at Scale},
    booktitle = {International Conference on Learning Representations},
    year      = {2021},
    eprint    = {2010.11929},
    archivePrefix = {arXiv}
}

@ARTICLE{emsellem1994,
  author  = {Emsellem, E. and Monnet, G. and Bacon, R.},
  title   = {The multi-gaussian expansion method: a tool for building
             realistic photometric and kinematical models of stellar
             systems. I. The formalism},
  journal = {A\&A},
  year    = {1994},
  volume  = {285},
  pages   = {723}
}

@ARTICLE{cappellari2002,
  author  = {Cappellari, Michele},
  title   = {Efficient multi-Gaussian expansion of galaxies},
  journal = {MNRAS},
  year    = {2002},
  volume  = {333},
  number  = {2},
  pages   = {400},
  doi     = {10.1046/j.1365-8711.2002.05412.x}
}

@ARTICLE{marti2014,
  author  = {Mart\'i, P. and Miquel, R. and Castander, F. J. and
             Gazta\~{n}aga, E. and Eriksen, M. and S\'anchez, C.},
  title   = {Precise photometric redshifts with a narrow-band filter set:
             the PAU survey at the William Herschel Telescope},
  journal = {MNRAS},
  year    = {2014},
  volume  = {442},
  number  = {1},
  pages   = {92},
  doi     = {10.1093/mnras/stu801}
}

@ARTICLE{eriksen2019,
  author  = {Eriksen, M. and Alarcon, A. and Gazta\~{n}aga, E. and others},
  title   = {The PAU Survey: early demonstration of photometric redshift
             performance in the COSMOS field},
  journal = {MNRAS},
  year    = {2019},
  volume  = {484},
  number  = {3},
  pages   = {4200},
  doi     = {10.1093/mnras/stz204}
}

@ARTICLE{serrano2023,
  author  = {Serrano, S. and Gazta\~{n}aga, E. and Castander, F. J. and others},
  title   = {The Physics of the Accelerating Universe Survey: narrow-band
             image photometry},
  journal = {MNRAS},
  year    = {2023},
  volume  = {523},
  number  = {3},
  pages   = {3287},
  doi     = {10.1093/mnras/stad1399},
  eprint  = {2206.14022}
}

@ARTICLE{bonoli2021,
  author  = {Bonoli, S. and Mar\'in-Franch, A. and Varela, J. and others},
  title   = {The miniJPAS survey: a preview of the Universe in 56 colours},
  journal = {A\&A},
  year    = {2021},
  volume  = {653},
  pages   = {A31},
  doi     = {10.1051/0004-6361/202038841}
}

@ARTICLE{mci2025,
  author  = {Zheng, Zhen-Ya and Xu, Chun and Liu, Xiaohua and Chen, Yong-He
             and Xu, Fang and Zhan, Hu and Li, Xinfeng and Zheng, Lixin
             and Shan, Huanyuan and Zhong, Jing and Yan, Zhaojun
             and Yuan, Fang-Ting and Jiang, Chunyan and Peng, Xiyan
             and Chen, Wei and Cheng, Xue and Chen, Zhen-Lei
             and Zhu, Shuairu and Long, Lin and Zhang, Xin and Gong, Yan
             and Shao, Li and Wang, Wei and Zhang, Tianyi and Ju, Guohao
             and Li, Chenghao and Wang, Wei and Li, Zhiyuan and Wang, Tao
             and Wang, Junfeng and Li, Chengyuan and Ma, Bin and Wang, Jianguo
             and Wang, Lei and Liu, Dezi and Lin, Nie and Li, Kexin
             and Wen, Xinrong and Wu, Maochun and Lin, Ruqiu and Ji, Xiang},
  title   = {{MCI}: Multi-channel Imager on the Chinese Space Station Survey
             Telescope},
  journal = {Research in Astronomy and Astrophysics},
  year    = {2026},
  volume  = {26},
  number  = {5},
  pages   = {055020},
  doi     = {10.1088/1674-4527/ae4a05},
  eprint  = {2509.14691},
  archivePrefix = {arXiv},
  primaryClass = {astro-ph.IM}
}

@ARTICLE{miyazaki2018,
  author  = {Miyazaki, Satoshi and Komiyama, Yutaka and Kawanomoto, Satoshi
             and others},
  title   = {Hyper Suprime-Cam: system design and verification of image
             quality},
  journal = {PASJ},
  year    = {2018},
  volume  = {70},
  pages   = {S1},
  doi     = {10.1093/pasj/psx063}
}

@ARTICLE{nightingale2021,
  author  = {Nightingale, James W. and Hayes, Richard G. and Kelly, Ashley
             and others},
  title   = {{PyAutoLens}: open-source strong gravitational lensing},
  journal = {JOSS},
  year    = {2021},
  volume  = {6},
  number  = {58},
  pages   = {2825},
  doi     = {10.21105/joss.02825}
}

@ARTICLE{sonnenfeld2018,
       author = {{Sonnenfeld}, Alessandro and {Chan}, James H. H. and
                 {Shu}, Yiping and {Wong}, Kenneth C. and
                 {Lee}, Chien-Hsiu and {Suyu}, Sherry H. and
                 {More}, Anupreeta and {Oguri}, Masamune and
                 {Price}, Paul and {Jaelani}, Anton T. and
                 {Kayo}, Issha and {Yasuda}, Naoki and
                 {Coupon}, Jean and {Tanaka}, Masayuki and
                 {Miyazaki}, Satoshi},
        title = "{Survey of Gravitationally-lensed Objects in HSC Imaging (SuGOHI). I. Automatic search for galaxy-scale strong lenses}",
      journal = {\pasj},
         year = 2018,
        month = jan,
       volume = {70},
       number = {SP1},
          eid = {S29},
        pages = {S29},
          doi = {10.1093/pasj/psx062},
archivePrefix = {arXiv},
       eprint = {1704.01585},
 primaryClass = {astro-ph.GA},
       adsurl = {https://ui.adsabs.harvard.edu/abs/2018PASJ...70S..29S}
}

@ARTICLE{petrillo2017,
       author = {{Petrillo}, C.~E. and {Tortora}, C. and
                 {Chatterjee}, S. and {Vernardos}, G. and
                 {Koopmans}, L.~V.~E. and {Verdoes Kleijn}, G. and
                 {Napolitano}, N.~R. and {Covone}, G. and
                 {Schneider}, P. and {Grado}, A. and
                 {McFarland}, J.},
        title = "{Finding strong gravitational lenses in the Kilo Degree Survey with Convolutional Neural Networks}",
      journal = {\mnras},
         year = 2017,
        month = nov,
       volume = {472},
       number = {1},
        pages = {1129-1150},
          doi = {10.1093/mnras/stx2052},
archivePrefix = {arXiv},
       eprint = {1702.07675},
 primaryClass = {astro-ph.GA},
       adsurl = {https://ui.adsabs.harvard.edu/abs/2017MNRAS.472.1129P}
}

@ARTICLE{wong2022,
       author = {{Wong}, Kenneth C. and {Chan}, James H. H. and
                 {Chao}, Dani C.-Y. and {Lee}, Chien-Hsiu and
                 {More}, Anupreeta and {Oguri}, Masamune},
        title = "{Survey of Gravitationally lensed objects in HSC Imaging (SuGOHI). VIII. New galaxy-scale lenses from the HSC SSP}",
      journal = {\pasj},
         year = 2022,
        month = oct,
       volume = {74},
       number = {5},
        pages = {1209-1219},
          doi = {10.1093/pasj/psac065},
       adsurl = {https://ui.adsabs.harvard.edu/abs/2022PASJ...74.1209W}
}

@ARTICLE{jaelani2024,
       author = {{Jaelani}, Anton T. and {More}, Anupreeta and
                 {Wong}, Kenneth C. and {Inoue}, Kaiki T. and
                 {Chao}, Dani C.-Y. and {Premadi}, Premana W. and
                 {Ca{\~n}ameras}, Raoul},
        title = "{Survey of gravitationally lensed objects in HSC imaging
                 (SuGOHI) -- X. Strong lens finding in the HSC-SSP using
                 convolutional neural networks}",
      journal = {\mnras},
         year = {2024},
        month = dec,
       volume = {535},
       number = {2},
        pages = {1625-1639},
          doi = {10.1093/mnras/stae2442}
}

@ARTICLE{grespan2024,
       author = {{Grespan}, Margherita and {Thuruthipilly}, Hareesh and
                 {Pollo}, Agnieszka and {Lochner}, Michelle and
                 {Biesiada}, Marek and {Etsebeth}, Verlon},
        title = "{{TEGLIE}: Transformer encoders as strong gravitational lens
                 finders in {KiDS}. From simulations to surveys}",
      journal = {\aap},
         year = {2024},
        month = aug,
       volume = {688},
          eid = {A34},
        pages = {A34},
          doi = {10.1051/0004-6361/202449929},
archivePrefix = {arXiv},
       eprint = {2405.11992},
 primaryClass = {astro-ph.GA}
}

@ARTICLE{gonzalez2026,
       author = {{Gonz{\'a}lez}, J. and {Holloway}, P. and
                 {Collett}, T. and {Verma}, A. and {Bechtol}, K. and
                 {Marshall}, P. and {More}, A. and others},
        title = "{Discovering Strong Gravitational Lenses in the Dark Energy
                 Survey with Interactive Machine Learning and Crowd-sourced
                 Inspection with Space Warps}",
      journal = {\apj},
         year = {2026},
        month = may,
       volume = {1002},
       number = {2},
          eid = {116},
        pages = {116},
          doi = {10.3847/1538-4357/ae450c},
archivePrefix = {arXiv},
       eprint = {2501.15679},
 primaryClass = {astro-ph.GA}
}

@ARTICLE{vincken2026,
  shortauthor = {Euclid Collaboration: Vincken {et~al.}},
       author = {{Euclid Collaboration} and {Vincken}, S.~H. and others},
        title = "{Euclid Quick Data Release (Q1). {AstroVink}: A vision
                 transformer approach to find strong gravitational lens
                 systems}",
      journal = {arXiv e-prints},
         year = {2026},
        month = apr,
          eid = {arXiv:2604.21977},
archivePrefix = {arXiv},
       eprint = {2604.21977},
 primaryClass = {astro-ph.IM},
          doi = {10.48550/arXiv.2604.21977}
}

@ARTICLE{walmsley2026,
  shortauthor = {Euclid Collaboration: Walmsley {et~al.}},
       author = {{Euclid Collaboration} and {Walmsley}, M. and
                 {Holloway}, P. and {Lines}, N.~E.~P. and
                 {Rojas}, K. and {Collett}, T.~E. and
                 {Verma}, A. and {Li}, T. and
                 {Nightingale}, J.~W. and {Despali}, G. and others},
        title = "{Euclid Quick Data Release (Q1): XXVI. The Strong Lensing Discovery Engine A -- System overview and first lens sample}",
      journal = {\aap},
         year = 2026,
        month = jul,
       volume = {711},
          eid = {A26},
        pages = {A26},
          doi = {10.1051/0004-6361/202555141},
archivePrefix = {arXiv},
       eprint = {2503.15324},
 primaryClass = {astro-ph.GA},
       adsurl = {https://ui.adsabs.harvard.edu/abs/2026A&A...711A..26E}
}

@ARTICLE{salvato2019,
       author = {{Salvato}, Mara and {Ilbert}, Olivier and {Hoyle}, Ben},
        title = "{The many flavours of photometric redshifts}",
      journal = {Nature Astronomy},
         year = 2019,
        month = mar,
       volume = {3},
        pages = {212-222},
          doi = {10.1038/s41550-018-0478-0},
       adsurl = {https://ui.adsabs.harvard.edu/abs/2019NatAs...3..212S}
}

@ARTICLE{ilbert2006,
       author = {{Ilbert}, O. and {Arnouts}, S. and {McCracken}, H.~J. and
                 {Bolzonella}, M. and {Bertin}, E. and {Le F{\`e}vre}, O. and
                 {Mellier}, Y. and {Zamorani}, G. and {Pell{\'o}}, R. and
                 {Iovino}, A. and others},
        title = "{Accurate photometric redshifts for the CFHT Legacy Survey calibrated using the VIMOS VLT Deep Survey}",
      journal = {\aap},
         year = 2006,
        month = oct,
       volume = {457},
        pages = {841-856},
          doi = {10.1051/0004-6361:20065138},
archivePrefix = {arXiv},
       eprint = {astro-ph/0603217},
       adsurl = {https://ui.adsabs.harvard.edu/abs/2006A&A...457..841I}
}

@ARTICLE{langeroodi2023,
       author = {{Langeroodi}, Danial and {Sonnenfeld}, Alessandro and
                 {Hoekstra}, Henk and {Agnello}, Adriano},
        title = "{Photometric redshift estimation of strongly lensed galaxies}",
      journal = {\aap},
         year = 2023,
        month = jan,
       volume = {669},
          eid = {A154},
        pages = {A154},
          doi = {10.1051/0004-6361/202244370},
archivePrefix = {arXiv},
       eprint = {2207.00595},
 primaryClass = {astro-ph.GA},
       adsurl = {https://ui.adsabs.harvard.edu/abs/2023A&A...669A.154L}
}

@ARTICLE{sonnenfeld2022,
       author = {{Sonnenfeld}, Alessandro},
        title = "{Statistical strong lensing. IV. Inferences with no individual source redshifts}",
      journal = {\aap},
         year = 2022,
       volume = {659},
          eid = {A133},
        pages = {A133},
          doi = {10.1051/0004-6361/202142467},
archivePrefix = {arXiv},
       eprint = {2110.09537},
 primaryClass = {astro-ph.GA},
       adsurl = {https://ui.adsabs.harvard.edu/abs/2022A&A...659A.133S}
}

@ARTICLE{alarcon2021,
       author = {{Alarcon}, Alex and {Gaztanaga}, Enrique and
                 {Eriksen}, Martin and {Baugh}, Carlton M. and
                 {Cabayol}, Laura and {Casas}, Ricard and
                 {Carretero}, Jorge and {Castander}, Francisco J. and
                 {De Vicente}, Juan and {Fernandez}, Enrique and others},
        title = "{The PAU Survey: an improved photo-z sample in the COSMOS field}",
      journal = {\mnras},
         year = 2021,
        month = mar,
       volume = {501},
       number = {4},
        pages = {6103-6122},
          doi = {10.1093/mnras/staa3659},
archivePrefix = {arXiv},
       eprint = {2007.11132},
 primaryClass = {astro-ph.GA},
       adsurl = {https://ui.adsabs.harvard.edu/abs/2021MNRAS.501.6103A}
}

@ARTICLE{benitez2014,
       author = {{Ben{\'i}tez}, N. and {Dupke}, R. and {Moles}, M. and
                 {Sodr{\'e}}, L. and {Cenarro}, J. and {Mar{\'i}n-Franch}, A. and
                 {Taylor}, K. and {Crist{\'o}bal}, D. and
                 {Fern{\'a}ndez-Soto}, A. and {Mendes de Oliveira}, C. and others},
        title = "{J-PAS: The Javalambre-Physics of the Accelerated Universe Astrophysical Survey}",
      journal = {arXiv e-prints},
         year = 2014,
        month = mar,
          eid = {arXiv:1403.5237},
archivePrefix = {arXiv},
       eprint = {1403.5237},
 primaryClass = {astro-ph.CO},
       adsurl = {https://ui.adsabs.harvard.edu/abs/2014arXiv1403.5237B}
}

@misc{yin2025,
    author        = {Yin, B. and others},
    title         = {{Dark Energy Survey Year 6 Results:
                     Redshift Calibration of the Weak Lensing
                     Source Galaxies}},
    year          = {2025},
    eprint        = {2510.23566},
    archivePrefix = {arXiv},
    primaryClass  = {astro-ph.CO},
    doi           = {10.48550/arXiv.2510.23566}
}

@ARTICLE{conselice2000,
       author = {{Conselice}, Christopher J. and
                 {Bershady}, Matthew A. and
                 {Jangren}, Anna},
        title = "{The Asymmetry of Galaxies: Physical Morphology for Nearby
                  and High-Redshift Galaxies}",
      journal = {\apj},
         year = 2000,
        month = feb,
       volume = {529},
       number = {2},
        pages = {886-910},
          doi = {10.1086/308300},
archivePrefix = {arXiv},
       eprint = {astro-ph/9907399},
       adsurl = {https://ui.adsabs.harvard.edu/abs/2000ApJ...529..886C}
}

@ARTICLE{conselice2003,
       author = {{Conselice}, Christopher J.},
        title = "{The Relationship between Stellar Light Distributions of
                  Galaxies and Their Formation Histories}",
      journal = {\apjs},
         year = 2003,
        month = jul,
       volume = {147},
       number = {1},
        pages = {1-28},
          doi = {10.1086/375001},
archivePrefix = {arXiv},
       eprint = {astro-ph/0303065},
       adsurl = {https://ui.adsabs.harvard.edu/abs/2003ApJS..147....1C}
}

@ARTICLE{gwyn2012,
       author = {{Gwyn}, Stephen D.~J.},
        title = "{The Canada-France-Hawaii Telescope Legacy Survey:
                  Stacked Images and Catalogs}",
      journal = {\aj},
         year = 2012,
        month = feb,
       volume = {143},
       number = {2},
          eid = {38},
        pages = {38},
          doi = {10.1088/0004-6256/143/2/38},
archivePrefix = {arXiv},
       eprint = {1101.1084},
 primaryClass = {astro-ph.CO},
       adsurl = {https://ui.adsabs.harvard.edu/abs/2012AJ....143...38G}
}

@ARTICLE{aihara2018,
       author = {{Aihara}, Hiroaki and {Arimoto}, Nobuo and
                 {Armstrong}, Robert and {Arnouts}, St{\'e}phane and
                 {Bahcall}, Neta A. and {Bickerton}, Steven and
                 {Bosch}, James and {Bundy}, Kevin and others},
        title = "{The Hyper Suprime-Cam SSP Survey: Overview and survey design}",
      journal = {\pasj},
         year = 2018,
        month = jan,
       volume = {70},
       number = {SP1},
          eid = {S4},
        pages = {S4},
          doi = {10.1093/pasj/psx066},
archivePrefix = {arXiv},
       eprint = {1704.05858},
 primaryClass = {astro-ph.IM},
       adsurl = {https://ui.adsabs.harvard.edu/abs/2018PASJ...70S...4A}
}

@ARTICLE{ivezic2019,
       author = {{Ivezi{\'c}}, {\v Z}eljko and {Kahn}, Steven M. and
                 {Tyson}, J. Anthony and {Abel}, Bob and
                 {Acosta}, Emily and {Allsman}, Robyn and
                 {Alonso}, David and {AlSayyad}, Yusra and others},
        title = "{LSST: From Science Drivers to Reference Design and
                  Anticipated Data Products}",
      journal = {\apj},
         year = 2019,
        month = mar,
       volume = {873},
       number = {2},
          eid = {111},
        pages = {111},
          doi = {10.3847/1538-4357/ab042c},
archivePrefix = {arXiv},
       eprint = {0805.2366},
 primaryClass = {astro-ph},
       adsurl = {https://ui.adsabs.harvard.edu/abs/2019ApJ...873..111I}
}

@article{cabanac2007,
  author  = {Cabanac, R. A. and others},
  title   = {The {CFHTLS} strong lensing legacy survey. I. Survey overview and {T0002} release sample},
  journal = {Astronomy and Astrophysics},
  year    = {2007},
  volume  = {461},
  pages   = {813--821},
  doi     = {10.1051/0004-6361:20065810},
  adsurl  = {https://ui.adsabs.harvard.edu/abs/2007A&A...461..813C/abstract}
}

@article{more2012,
  author  = {More, A. and others},
  title   = {The {CFHTLS}-Strong Lensing Legacy Survey ({SL2S}): Investigating the Group-scale Lenses with the {SARCS} Sample},
  journal = {The Astrophysical Journal},
  year    = {2012},
  volume  = {749},
  eid     = {38},
  doi     = {10.1088/0004-637X/749/1/38},
  adsurl  = {https://ui.adsabs.harvard.edu/abs/2012ApJ...749...38M/abstract}
}

@article{sonnenfeld2013,
  author  = {Sonnenfeld, Alessandro and others},
  title   = {The {SL2S} Galaxy-scale Lens Sample. III. Lens Models, Surface Photometry, and Stellar Masses for the Final Sample},
  journal = {The Astrophysical Journal},
  year    = {2013},
  volume  = {777},
  eid     = {97},
  doi     = {10.1088/0004-637X/777/2/97},
  adsurl  = {https://ui.adsabs.harvard.edu/abs/2013ApJ...777...97S/abstract}
}

@article{gavazzi2014,
  author  = {Gavazzi, Rapha{\"e}l and others},
  title   = {{RINGFINDER}: Automated Detection of Galaxy-scale Gravitational Lenses in Ground-based Multi-filter Imaging Data},
  journal = {The Astrophysical Journal},
  year    = {2014},
  volume  = {785},
  eid     = {144},
  doi     = {10.1088/0004-637X/785/2/144},
  adsurl  = {https://ui.adsabs.harvard.edu/abs/2014ApJ...785..144G/abstract}
}

@article{garvin2022,
  author  = {Garvin, Emily O. and others},
  title   = {Hubble Asteroid Hunter. II. Identifying Strong Gravitational Lenses in {HST} Images with Crowdsourcing},
  journal = {Astronomy and Astrophysics},
  year    = {2022},
  volume  = {667},
  eid     = {A141},
  doi     = {10.1051/0004-6361/202243745},
  adsurl  = {https://ui.adsabs.harvard.edu/abs/2022A&A...667A.141G/abstract}
}

@article{chan2020,
  author  = {Chan, James H. H. and others},
  title   = {Survey of Gravitationally-lensed Objects in {HSC} Imaging ({SuGOHI}). IV. Lensed Quasar Search in the {HSC} Survey},
  journal = {Astronomy and Astrophysics},
  year    = {2020},
  volume  = {636},
  eid     = {A87},
  doi     = {10.1051/0004-6361/201937030},
  adsurl  = {https://ui.adsabs.harvard.edu/abs/2020A&A...636A..87C/abstract}
}

@article{negrello2010,
  author  = {Negrello, Mattia and others},
  title   = {The Detection of a Population of Submillimeter-Bright, Strongly Lensed Galaxies},
  journal = {Science},
  year    = {2010},
  volume  = {330},
  pages   = {800--804},
  doi     = {10.1126/science.1193420},
  adsurl  = {https://ui.adsabs.harvard.edu/abs/2010Sci...330..800N/abstract}
}

@article{spiniello2018,
  author  = {Spiniello, C. and others},
  title   = {{KiDS-SQuaD}: The {KiDS} Strongly Lensed Quasar Detection Project},
  journal = {Monthly Notices of the Royal Astronomical Society},
  year    = {2018},
  volume  = {480},
  pages   = {1163--1173},
  doi     = {10.1093/mnras/sty1923},
  adsurl  = {https://ui.adsabs.harvard.edu/abs/2018MNRAS.480.1163S/abstract}
}

@article{kronemartins2019,
  author        = {Krone-Martins, A. and others},
  title         = {{Gaia GraL}: {Gaia DR2} Gravitational Lens Systems. V. Doubly-imaged {QSOs} Discovered from Entropy and Wavelets},
  journal       = {arXiv e-prints},
  year          = {2019},
  eid           = {arXiv:1912.08977},
  archivePrefix = {arXiv},
  eprint        = {1912.08977},
  adsurl        = {https://ui.adsabs.harvard.edu/abs/2019arXiv191208977K/abstract}
}

@article{pourrahmani2018,
  author  = {Pourrahmani, Milad and others},
  title   = {{LensFlow}: A Convolutional Neural Network in Search of Strong Gravitational Lenses},
  journal = {The Astrophysical Journal},
  year    = {2018},
  volume  = {856},
  eid     = {68},
  doi     = {10.3847/1538-4357/aaae6a},
  adsurl  = {https://ui.adsabs.harvard.edu/abs/2018ApJ...856...68P/abstract}
}

@article{jackson2008,
  author  = {Jackson, Neal},
  title   = {Gravitational Lenses and Lens Candidates Identified from the {COSMOS} Field},
  journal = {Monthly Notices of the Royal Astronomical Society},
  year    = {2008},
  volume  = {389},
  pages   = {1311--1318},
  doi     = {10.1111/j.1365-2966.2008.13629.x},
  adsurl  = {https://ui.adsabs.harvard.edu/abs/2008MNRAS.389.1311J/abstract}
}

@article{vanderwel2013,
  author  = {van der Wel, A. and others},
  title   = {Discovery of a Quadruple Lens in {CANDELS} with a Record Lens Redshift $z=1.53$},
  journal = {The Astrophysical Journal Letters},
  year    = {2013},
  volume  = {777},
  eid     = {L17},
  doi     = {10.1088/2041-8205/777/1/L17},
  adsurl  = {https://ui.adsabs.harvard.edu/abs/2013ApJ...777L..17V/abstract}
}

@article{fischer1998,
  author  = {Fischer, Philippe and others},
  title   = {Discovery of a New Quadruple Lens {HST 1411+5211}},
  journal = {The Astrophysical Journal Letters},
  year    = {1998},
  volume  = {503},
  pages   = {L127--L130},
  doi     = {10.1086/311555},
  adsurl  = {https://ui.adsabs.harvard.edu/abs/1998ApJ...503L.127F/abstract}
}

@article{more2016,
  author  = {More, Anupreeta and others},
  title   = {The {SDSS-III BOSS} Quasar Lens Survey: Discovery of 13 Gravitationally Lensed Quasars},
  journal = {Monthly Notices of the Royal Astronomical Society},
  year    = {2016},
  volume  = {456},
  pages   = {1595--1606},
  doi     = {10.1093/mnras/stv2813},
  adsurl  = {https://ui.adsabs.harvard.edu/abs/2016MNRAS.456.1595M/abstract}
}

@article{bolton2004,
  author  = {Bolton, Adam S. and others},
  title   = {Sloan Digital Sky Survey Spectroscopic Lens Search. I. Discovery of Intermediate-Redshift Star-forming Galaxies behind Foreground Luminous Red Galaxies},
  journal = {The Astronomical Journal},
  year    = {2004},
  volume  = {127},
  pages   = {1860--1882},
  doi     = {10.1086/382714},
  adsurl  = {https://ui.adsabs.harvard.edu/abs/2004AJ....127.1860B/abstract}
}

@article{ratnatunga1995,
  author  = {Ratnatunga, Kavan U. and others},
  title   = {New ``Einstein Cross'' Gravitational Lens Candidates in {Hubble Space Telescope WFPC2} Survey Images},
  journal = {The Astrophysical Journal Letters},
  year    = {1995},
  volume  = {453},
  pages   = {L5--L8},
  doi     = {10.1086/309738},
  adsurl  = {https://ui.adsabs.harvard.edu/abs/1995ApJ...453L...5R/abstract}
}

@article{moustakas2007,
  author  = {Moustakas, Leonidas A. and others},
  title   = {A Strong-Lens Survey in {AEGIS}: The Influence of Large-Scale Structure},
  journal = {The Astrophysical Journal Letters},
  year    = {2007},
  volume  = {660},
  pages   = {L31--L34},
  doi     = {10.1086/517930},
  adsurl  = {https://ui.adsabs.harvard.edu/abs/2007ApJ...660L..31M/abstract}
}

@article{jaelani2021,
  author  = {Jaelani, Anton T. and others},
  title   = {Survey of Gravitationally Lensed Objects in {HSC} Imaging ({SuGOHI}). VII. Discovery and Confirmation of Three Strongly Lensed Quasars},
  journal = {Monthly Notices of the Royal Astronomical Society},
  year    = {2021},
  volume  = {502},
  pages   = {1487--1493},
  doi     = {10.1093/mnras/stab145},
  adsurl  = {https://ui.adsabs.harvard.edu/abs/2021MNRAS.502.1487J/abstract}
}

@article{sand2005,
  author  = {Sand, David J. and others},
  title   = {A Systematic Search for Gravitationally Lensed Arcs in the {Hubble Space Telescope WFPC2} Archive},
  journal = {The Astrophysical Journal},
  year    = {2005},
  volume  = {627},
  pages   = {32--52},
  doi     = {10.1086/430298},
  adsurl  = {https://ui.adsabs.harvard.edu/abs/2005ApJ...627...32S/abstract}
}

@ARTICLE{sonnenfeld2019,
       author = {{Sonnenfeld}, Alessandro and {Jaelani}, Anton T. and
                 {Chan}, James H. H. and {More}, Anupreeta and
                 {Suyu}, Sherry H. and {Wong}, Kenneth C. and
                 {Oguri}, Masamune and {Lee}, Chien-Hsiu},
        title = "{Survey of gravitationally-lensed objects in HSC imaging
                 (SuGOHI). III. Statistical strong lensing constraints on the
                 stellar IMF of CMASS galaxies}",
      journal = {\aap},
         year = {2019},
        month = oct,
       volume = {630},
          eid = {A71},
        pages = {A71},
          doi = {10.1051/0004-6361/201935743}
}

@ARTICLE{dye2014,
       author = {{Dye}, S. and {Negrello}, M. and {Hopwood}, R. and
                 {Nightingale}, J. W. and {Bussmann}, R. S. and others},
        title = "{Herschel-ATLAS: modelling the first strong gravitational
                 lenses}",
      journal = {\mnras},
         year = {2014},
        month = may,
       volume = {440},
       number = {3},
        pages = {2013-2025},
          doi = {10.1093/mnras/stu305},
archivePrefix = {arXiv},
       eprint = {1311.5893},
 primaryClass = {astro-ph.CO}
}

@ARTICLE{wong2017,
       author = {{Wong}, Kenneth C. and {Ishida}, Tsuyoshi and
                 {Tamura}, Yoichi and {Suyu}, Sherry H. and
                 {Oguri}, Masamune and {Matsushita}, Satoki},
        title = "{ALMA Observations of the Gravitational Lens SDP.9}",
      journal = {\apjl},
         year = {2017},
        month = jul,
       volume = {843},
       number = {2},
          eid = {L35},
        pages = {L35},
          doi = {10.3847/2041-8213/aa7d4a},
archivePrefix = {arXiv},
       eprint = {1707.00702},
 primaryClass = {astro-ph.GA}
}

@ARTICLE{bolton2008,
       author = {{Bolton}, Adam S. and {Burles}, Scott and
                 {Koopmans}, Leon V. E. and {Treu}, Tommaso and
                 {Gavazzi}, Rapha{\"e}l and {Moustakas}, Leonidas A. and
                 {Wayth}, Randall and {Schlegel}, David J. and others},
        title = "{The Sloan Lens ACS Survey. V. The Full ACS Strong-Lens
                 Sample}",
      journal = {\apj},
         year = {2008},
        month = aug,
       volume = {682},
       number = {2},
        pages = {964-984},
          doi = {10.1086/589327},
archivePrefix = {arXiv},
       eprint = {0805.1931},
 primaryClass = {astro-ph}
}

@ARTICLE{eriksen2020,
       author = {{Eriksen}, M. and {Alarcon}, A. and {Cabayol}, L. and
                 {Carretero}, J. and {Casas}, R. and {Castander}, F. J. and
                 {De Vicente}, J. and {Fernandez}, E. and
                 {Garcia-Bellido}, J. and {Gaztanaga}, E. and others},
        title = "{The PAU Survey: Photometric redshifts using transfer learning
                 from simulations}",
      journal = {\mnras},
         year = {2020},
        month = oct,
       volume = {497},
       number = {4},
        pages = {4565-4579},
          doi = {10.1093/mnras/staa2265},
archivePrefix = {arXiv},
       eprint = {2004.07979},
 primaryClass = {astro-ph.GA}
}

@ARTICLE{bertin1996,
       author = {{Bertin}, E. and {Arnouts}, S.},
        title = "{SExtractor: Software for source extraction}",
      journal = {\aaps},
         year = {1996},
        month = jun,
       volume = {117},
        pages = {393-404},
          doi = {10.1051/aas:1996164}
}

@ARTICLE{navarroGirones2024,
       author = {{Navarro-Giron{\'e}s}, D. and {Gazta{\~n}aga}, E. and
                 {Crocce}, M. and {Wittje}, A. and {Hildebrandt}, H. and
                 {Wright}, A. H. and {Siudek}, M. and {Eriksen}, M. and
                 {Serrano}, S. and {Renard}, P. and others},
        title = "{The PAU survey: photometric redshift estimation in deep wide fields}",
      journal = {\mnras},
         year = {2024},
        month = oct,
       volume = {534},
       number = {2},
        pages = {1504-1527},
          doi = {10.1093/mnras/stae1686},
archivePrefix = {arXiv},
       eprint = {2312.07581},
 primaryClass = {astro-ph.CO}
}

@ARTICLE{wittman2016,
       author = {{Wittman}, David and {Bhaskar}, Ramya and {Tobin}, Ryan},
        title = "{Overconfidence in photometric redshift estimation}",
      journal = {\mnras},
         year = {2016},
       volume = {457},
       number = {4},
        pages = {4005-4011},
          doi = {10.1093/mnras/stw261},
archivePrefix = {arXiv},
       eprint = {1601.07857},
 primaryClass = {astro-ph.CO}
}

@ARTICLE{jones2019,
       author = {{Jones}, Daniel M. and {Heavens}, Alan F.},
        title = "{Bayesian photometric redshifts of blended sources}",
      journal = {\mnras},
         year = {2019},
       volume = {483},
       number = {2},
        pages = {2487-2505},
          doi = {10.1093/mnras/sty3279},
archivePrefix = {arXiv},
       eprint = {1808.02846},
 primaryClass = {astro-ph.CO}
}

@ARTICLE{metcalf2019,
       author = {{Metcalf}, R. B. and {Meneghetti}, M. and {Avestruz}, C. and
                 {Bellagamba}, F. and {Bom}, C. R. and {Bertin}, E. and
                 {Cabanac}, R. and {Courbin}, F. and {Davies}, A. and others},
        title = "{The strong gravitational lens finding challenge}",
      journal = {\aap},
         year = {2019},
       volume = {625},
          eid = {A119},
        pages = {A119},
          doi = {10.1051/0004-6361/201832797},
archivePrefix = {arXiv},
       eprint = {1802.03609},
 primaryClass = {astro-ph.GA}
}

@ARTICLE{hernanCaballero2024,
       author = {{Hern{\'a}n-Caballero}, A. and {Akhlaghi}, M. and
                 {L{\'o}pez-Sanjuan}, C. and {V{\'a}zquez-Rami{\'o}}, H. and
                 {Laur}, J. and {Varela}, J. and {Civera}, T. and
                 {Muniesa}, D. and {Finoguenov}, A. and
                 {Fern{\'a}ndez-Ontiveros}, J. A. and others},
        title = "{The miniJPAS survey: Maximising the photo-z accuracy from
                  multi-survey datasets with probability conflation}",
      journal = {\aap},
         year = {2024},
       volume = {684},
          eid = {A61},
        pages = {A61},
          doi = {10.1051/0004-6361/202348513},
archivePrefix = {arXiv},
       eprint = {2311.04220},
 primaryClass = {astro-ph.GA}
}

@ARTICLE{merlin2015,
       author = {{Merlin}, E. and {Fontana}, A. and {Ferguson}, H. C. and
                 {Dunlop}, J. S. and {Elbaz}, D. and {Bourne}, N. and
                 {Bruce}, V. A. and {Buitrago}, F. and {Castellano}, M. and
                 {Schreiber}, C. and others},
        title = "{T-PHOT: A new code for PSF-matched, prior-based,
                  multiwavelength extragalactic deconfusion photometry}",
      journal = {\aap},
         year = {2015},
       volume = {582},
          eid = {A15},
        pages = {A15},
          doi = {10.1051/0004-6361/201526471},
archivePrefix = {arXiv},
       eprint = {1505.02516},
 primaryClass = {astro-ph.IM}
}

@ARTICLE{euclidTarsitano2026,
  shortauthor = {Euclid Collaboration: Tarsitano {et~al.}},
       author = {{Euclid Collaboration} and {Tarsitano}, F. and
                 {Schreiber}, C. and {Miyatake}, H. and {Nishizawa}, A. J.
                 and {Hartley}, W. G. and {Miller}, L. and others},
        title = "{Euclid: IX. Galaxy SED reconstruction in the PHZ processing
                  function: Impact on the PSF and the role of medium-band filters}",
      journal = {\aap},
         year = {2026},
       volume = {713},
        pages = {A178},
          doi = {10.1051/0004-6361/202659032},
archivePrefix = {arXiv},
       eprint = {2601.10795},
 primaryClass = {astro-ph.CO}
}

@ARTICLE{bosch2018,
       author = {{Bosch}, James and {Armstrong}, Robert and {Bickerton}, Steven
                 and {Furusawa}, Hisanori and {Ikeda}, Hiroyuki and
                 {Koike}, Michitaro and {Lupton}, Robert and others},
        title = "{The Hyper Suprime-Cam software pipeline}",
      journal = {\pasj},
         year = {2018},
       volume = {70},
       number = {SP1},
        pages = {S5},
          doi = {10.1093/pasj/psx080},
archivePrefix = {arXiv},
       eprint = {1705.06766},
 primaryClass = {astro-ph.IM}
}

@ARTICLE{brammer2008,
       author = {{Brammer}, Gabriel B. and {van Dokkum}, Pieter G. and {Coppi}, Paolo},
        title = "{EAZY: A Fast, Public Photometric Redshift Code}",
      journal = {ApJ},
         year = 2008,
       volume = {686},
        pages = {1503-1513},
          doi = {10.1086/591786}
}

@ARTICLE{benitez2000,
       author = {{Ben{\'\i}tez}, Narciso},
        title = "{Bayesian Photometric Redshift Estimation}",
      journal = {ApJ},
         year = 2000,
       volume = {536},
        pages = {571-583},
          doi = {10.1086/308947}
}

@ARTICLE{arnouts1999,
       author = {{Arnouts}, S. and {Cristiani}, S. and {Moscardini}, L. and {Matarrese}, S. and {Lucchin}, F. and {Fontana}, A. and {Giallongo}, E.},
        title = "{Measuring and modelling the redshift evolution of clustering: the Hubble Deep Field North}",
      journal = {MNRAS},
         year = 1999,
       volume = {310},
        pages = {540-556},
          doi = {10.1046/j.1365-8711.1999.02978.x}
}

@ARTICLE{tanaka2018,
       author = {{Tanaka}, Masayuki and {Coupon}, Jean and {Hsieh}, Bau-Ching and others},
        title = "{Photometric redshifts for Hyper Suprime-Cam Subaru Strategic Program Data Release 1}",
      journal = {PASJ},
         year = 2018,
       volume = {70},
        pages = {S9},
          doi = {10.1093/pasj/psx077}
}

@ARTICLE{hildebrandt2012,
       author = {{Hildebrandt}, H. and {Erben}, T. and {Kuijken}, K. and others},
        title = "{CFHTLenS: improving the quality of photometric redshifts with precision photometry}",
      journal = {MNRAS},
         year = 2012,
       volume = {421},
        pages = {2355-2367},
          doi = {10.1111/j.1365-2966.2012.20468.x}
}

@ARTICLE{conroy2009,
       author = {{Conroy}, Charlie and {Gunn}, James E. and {White}, Martin},
        title = "{The Propagation of Uncertainties in Stellar Population Synthesis Modeling. I. The Relevance of Uncertain Aspects of Stellar Evolution and the Initial Mass Function to the Derived Physical Properties of Galaxies}",
      journal = {ApJ},
         year = 2009,
       volume = {699},
        pages = {486-506},
          doi = {10.1088/0004-637X/699/1/486}
}

@ARTICLE{vegetti2012,
  author = {{Vegetti}, S. and {Lagattuta}, D. J. and {McKean}, J. P. and
            {Auger}, M. W. and {Fassnacht}, C. D. and {Koopmans}, L. V. E.},
  title = {Gravitational detection of a low-mass dark satellite galaxy at cosmological distance},
  journal = {Nature},
  year = {2012},
  volume = {481},
  pages = {341--343},
  doi = {10.1038/nature10669},
  archivePrefix = {arXiv},
  eprint = {1201.3643}
}

@ARTICLE{gilman2020,
  author = {{Gilman}, Daniel and {Birrer}, Simon and {Nierenberg}, Anna and
            {Treu}, Tommaso and {Du}, Xiaolong and {Benson}, Andrew},
  title = {Warm dark matter chills out: constraints on the halo mass function and the free-streaming length of dark matter with eight quadruple-image strong gravitational lenses},
  journal = {\mnras},
  year = {2020},
  volume = {491},
  number = {4},
  pages = {6077--6101},
  doi = {10.1093/mnras/stz3480}
}

@ARTICLE{nightingale2024,
  author = {{Nightingale}, James W. and {He}, Qiuhan and {Cao}, Xiaoyue and
            {Amvrosiadis}, Aristeidis and {Etherington}, Amy and {Frenk}, Carlos S. and
            {Hayes}, Richard G. and {Robertson}, Andrew and {Cole}, Shaun and
            {Lange}, Samuel and {Li}, Ran and {Massey}, Richard},
  title = {Scanning for dark matter subhaloes in {Hubble Space Telescope} imaging of 54 strong lenses},
  journal = {\mnras},
  year = {2024},
  volume = {527},
  number = {4},
  pages = {10480--10506},
  doi = {10.1093/mnras/stad3694},
  archivePrefix = {arXiv},
  eprint = {2209.10566}
}

@ARTICLE{he2022,
  author = {{He}, Qiuhan and {Robertson}, Andrew and {Nightingale}, James and
            {Cole}, Shaun and {Frenk}, Carlos S. and {Massey}, Richard and
            {Amvrosiadis}, Aristeidis and {Li}, Ran and {Cao}, Xiaoyue and
            {Etherington}, Amy},
  title = {A forward-modelling method to infer the dark matter particle mass from strong gravitational lenses},
  journal = {\mnras},
  year = {2022},
  volume = {511},
  number = {2},
  pages = {3046--3062},
  doi = {10.1093/mnras/stac191},
  archivePrefix = {arXiv},
  eprint = {2010.13221}
}

@ARTICLE{he2023subhalo,
  author = {{He}, Qiuhan and {Nightingale}, James and {Robertson}, Andrew and
            {Amvrosiadis}, Aristeidis and {Cole}, Shaun and {Frenk}, Carlos S. and
            {Massey}, Richard and {Li}, Ran and {Amorisco}, Nicola C. and
            {Metcalf}, R. Benton and {Cao}, Xiaoyue and {Etherington}, Amy},
  title = {Testing strong lensing subhalo detection with a cosmological simulation},
  journal = {\mnras},
  year = {2023},
  volume = {518},
  number = {1},
  pages = {220--239},
  doi = {10.1093/mnras/stac2779},
  archivePrefix = {arXiv},
  eprint = {2202.10191}
}

@ARTICLE{tdcosmo2025,
  author = {{TDCOSMO Collaboration} and {Birrer}, Simon and
            {Buckley-Geer}, Elizabeth J. and {Cappellari}, Michele and
            {Courbin}, Fr{\'e}d{\'e}ric and {Dux}, Fr{\'e}d{\'e}ric and
            {Fassnacht}, Christopher D. and {Frieman}, Joshua A. and others},
  shortauthor = {TDCOSMO Collaboration: Birrer {et~al.}},
  title = {{TDCOSMO} 2025: Cosmological constraints from strong lensing time delays},
  journal = {\aap},
  year = {2025},
  volume = {704},
  pages = {A63},
  doi = {10.1051/0004-6361/202555801},
  archivePrefix = {arXiv},
  eprint = {2506.03023}
}

@ARTICLE{collettAuger2014,
  author = {{Collett}, Thomas E. and {Auger}, Matthew W.},
  title = {Cosmological constraints from the double source plane lens {SDSSJ0946+1006}},
  journal = {\mnras},
  year = {2014},
  volume = {443},
  number = {2},
  pages = {969--976},
  doi = {10.1093/mnras/stu1190},
  archivePrefix = {arXiv},
  eprint = {1403.5278}
}

@ARTICLE{cao2012cosmology,
  author = {{Cao}, Shuo and {Pan}, Yu and {Biesiada}, Marek and
            {Godlowski}, Wlodzimierz and {Zhu}, Zong-Hong},
  title = {Constraints on cosmological models from strong gravitational lensing systems},
  journal = {JCAP},
  year = {2012},
  volume = {03},
  pages = {016},
  doi = {10.1088/1475-7516/2012/03/016},
  archivePrefix = {arXiv},
  eprint = {1105.6226}
}

@ARTICLE{cao2015cosmology,
  author = {{Cao}, Shuo and {Biesiada}, Marek and {Gavazzi}, Rapha{\"e}l and
            {Pi{\'o}rkowska}, Aleksandra and {Zhu}, Zong-Hong},
  title = {Cosmology with Strong-Lensing Systems},
  journal = {\apj},
  year = {2015},
  volume = {806},
  number = {2},
  pages = {185},
  doi = {10.1088/0004-637X/806/2/185},
  archivePrefix = {arXiv},
  eprint = {1509.07649}
}

@ARTICLE{padilla2019,
  author = {{Padilla}, C. and {Castander}, F. J. and {Alarc{\'o}n}, A. and others},
  title = {The Physics of the Accelerating Universe Camera},
  journal = {\aj},
  year = {2019},
  volume = {157},
  number = {6},
  pages = {246},
  doi = {10.3847/1538-3881/ab0412},
  archivePrefix = {arXiv},
  eprint = {1902.03623}
}

@ARTICLE{renard2022d4000,
  author = {{Renard}, P. and {Siudek}, M. and {Eriksen}, M. B. and others},
  title = {The {PAU} Survey: Measurements of the 4000 {\AA} spectral break with narrow-band photometry},
  journal = {\mnras},
  year = {2022},
  volume = {515},
  number = {1},
  pages = {146--166},
  doi = {10.1093/mnras/stac1730},
  archivePrefix = {arXiv},
  eprint = {2201.04411}
}

@ARTICLE{torralba2024lya,
  author = {{Torralba-Torregrosa}, A. and {Renard}, P. and {Spinoso}, D. and others},
  title = {The {PAU} Survey: The quasar {Ly$\alpha$} and {UV} luminosity functions at {$2.7 < z < 5.3$}},
  journal = {\aap},
  year = {2024},
  volume = {690},
  pages = {A388},
  doi = {10.1051/0004-6361/202451675},
  archivePrefix = {arXiv},
  eprint = {2407.19020}
}

@ARTICLE{collister2004,
  author = {{Collister}, A. A. and {Lahav}, O.},
  title = {{ANNz}: Estimating Photometric Redshifts Using Artificial Neural Networks},
  journal = {\pasp},
  year = {2004},
  volume = {116},
  number = {818},
  pages = {345--351},
  doi = {10.1086/383254},
  archivePrefix = {arXiv},
  eprint = {astro-ph/0311058}
}

@ARTICLE{sadeh2016,
  author = {{Sadeh}, I. and {Abdalla}, F. B. and {Lahav}, O.},
  title = {{ANNz2}: Photometric Redshift and Probability Distribution Function Estimation using Machine Learning},
  journal = {\pasp},
  year = {2016},
  volume = {128},
  number = {968},
  pages = {104502},
  doi = {10.1088/1538-3873/128/968/104502},
  archivePrefix = {arXiv},
  eprint = {1507.00490}
}

@ARTICLE{tessore2015,
  author = {{Tessore}, N. and {Metcalf}, R. B.},
  title = {The elliptical power law profile lens},
  journal = {\aap},
  year = {2015},
  volume = {580},
  pages = {A79},
  doi = {10.1051/0004-6361/201526773},
  archivePrefix = {arXiv},
  eprint = {1507.01819}
}

@ARTICLE{keeton1997,
  author = {{Keeton}, C. R. and {Kochanek}, C. S. and {Seljak}, U.},
  title = {Shear and Ellipticity in Gravitational Lenses},
  journal = {\apj},
  year = {1997},
  volume = {482},
  number = {2},
  pages = {604--620},
  doi = {10.1086/304172},
  archivePrefix = {arXiv},
  eprint = {astro-ph/9610163}
}

\begin{appendix}
\onecolumn
\section{Results for the remaining four systems}
\label{app:other-systems}

\begin{figure}[!ht]
\centering
\includegraphics[width=\textwidth]{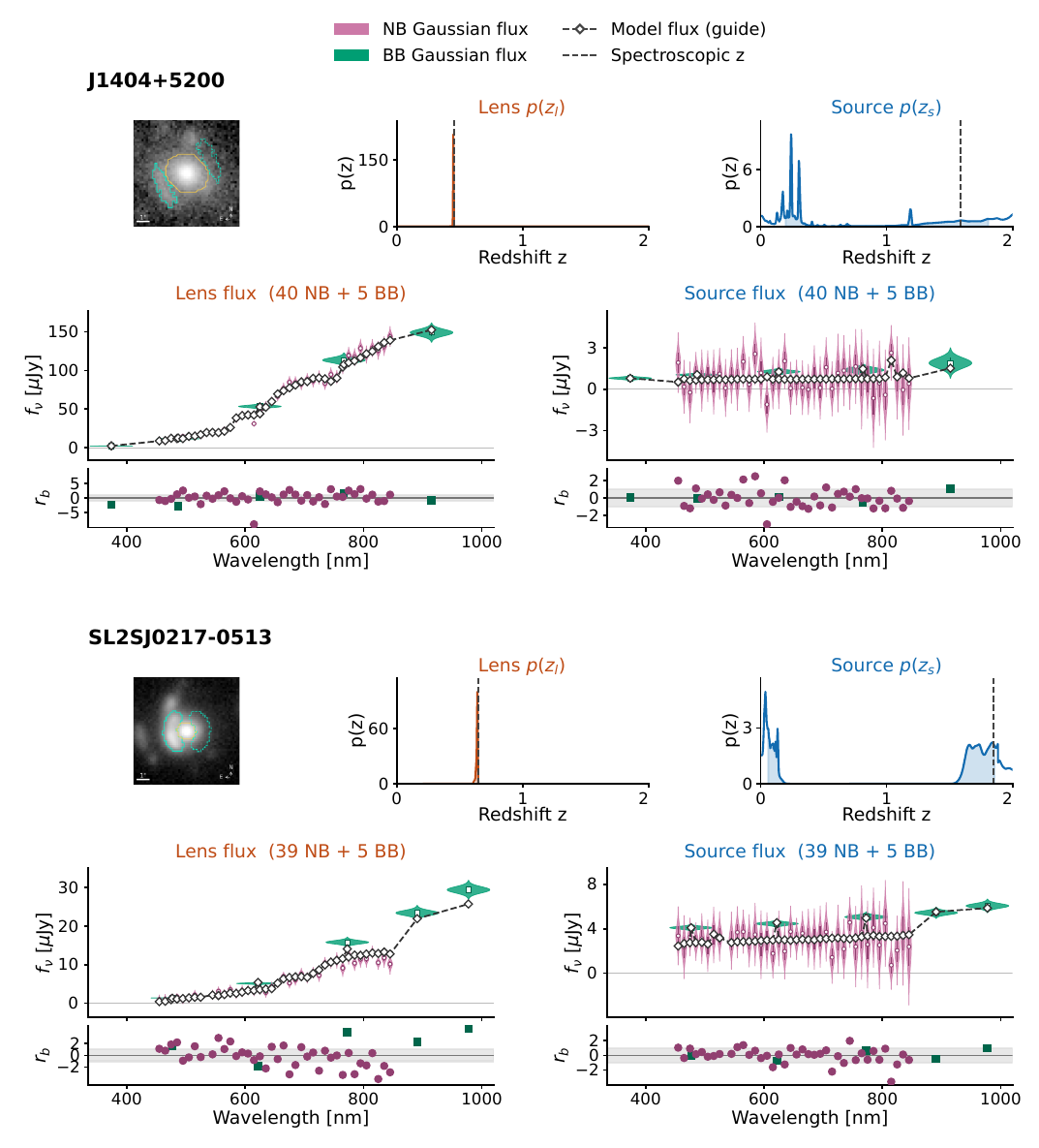}
\caption{Results for J1404$+$5200 and SL2SJ0217$-$0513. The flux, residual, and posterior layout is identical to Fig.~\ref{fig:results-a}. The panels show the source-peak offsets together with their broad redshift distributions.}
\label{fig:results-appendix-a}
\end{figure}

\begin{figure}[!ht]
\centering
\includegraphics[width=\textwidth]{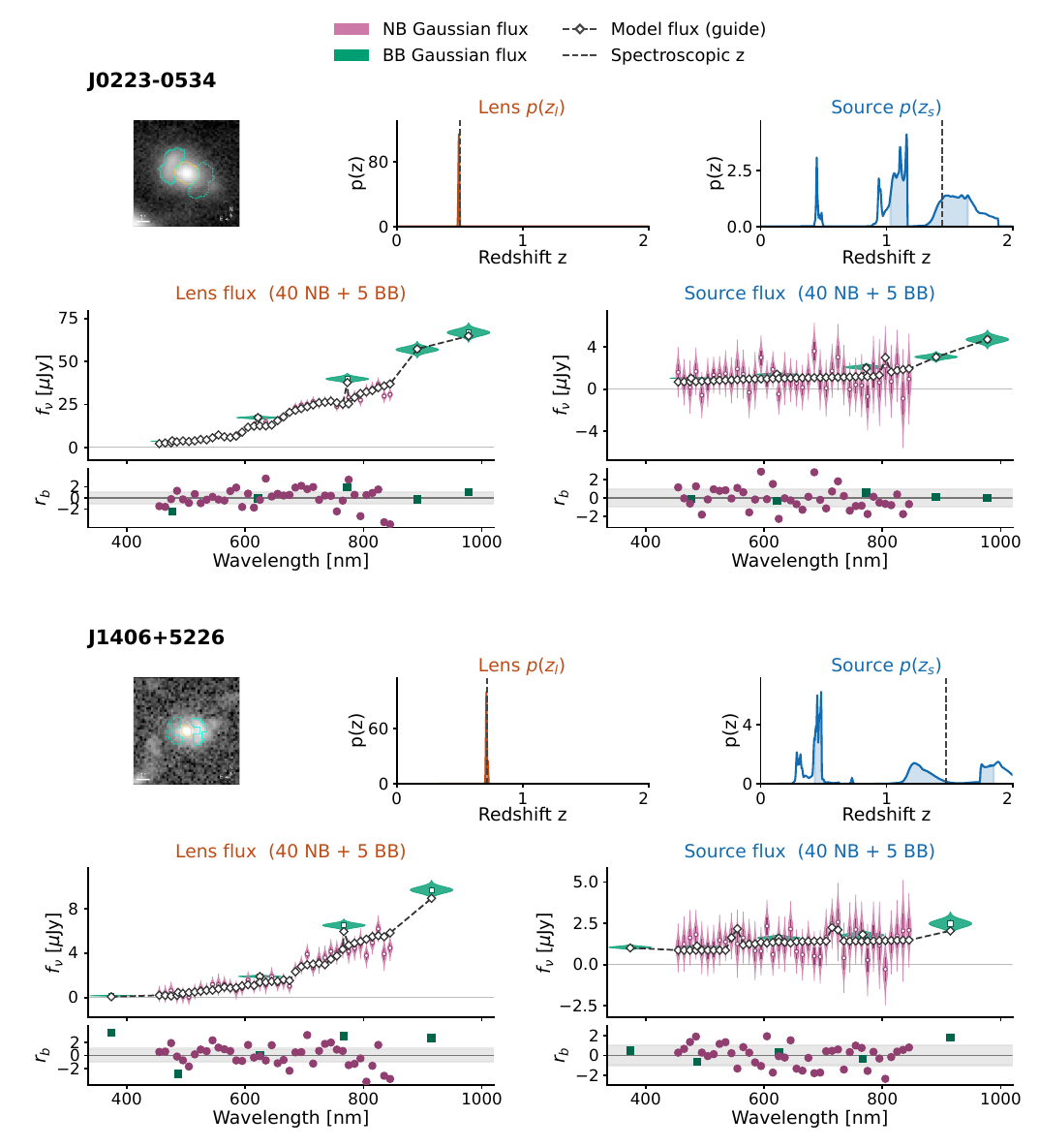}
\caption{Results for J0223$-$0534 and J1406$+$5226. J0223$-$0534 has a large ordering probability despite a source-peak discrepancy. J1406$+$5226 uses CFHT broad-band data.}
\label{fig:results-appendix-b}
\end{figure}

Figures~\ref{fig:results-appendix-a} and \ref{fig:results-appendix-b} show the four systems with source peaks outside $|\delta_{\rm s}|<0.05$. They use the same saved $2\sigma$-growth products and diagnostic layout as the main figures. All four remain in the main-text summary and both control comparisons.

\clearpage

\clearpage
\section{Reference imaging and reproducibility details}
\label{app:setup}
\begin{table}[ht]
\caption{Reference-band geometry for all eight systems. PSF widths are second-moment Gaussian-equivalent FWHM values used in region selection, and the actual PSF images are supplied. Areas refer to the native reference grid, with equal source and reflected-background areas. The centring annulus is specified in angular units and transferred to each PAUS image. The final column counts accepted offsets ($\leq0.15$ arcsec) among bands with centring data. A dash denotes unavailable centring radii. SDSSJ1416$+$5136 uses the broad-band centre in every band.}
\label{tab:setup}\centering\small
\setlength{\tabcolsep}{5pt}
\begin{tabular}{lrrrrrr}
\toprule
System & Reference & PSF FWHM & $A_{\rm L}$ & $A_{\rm S}=A_{\rm R}$ & $(r_{\rm in},r_{\rm out})$ & Shifted NB \\
 & band & [arcsec] & [arcsec$^2$] & [arcsec$^2$] & [arcsec] & bands \\
\midrule
SDSSJ0912$+$0029 & HSC G & 1.12 & 2.29 & 1.81 & (0.25, 1.45) & 23/26 \\
J0225$-$0454 & HSC G & 1.30 & 1.81 & 1.75 & (0.25, 0.76) & 35/40 \\
J1404$+$5200 & CFHT G & 1.36 & 8.99 & 3.22 & (0.28, 1.60) & 31/40 \\
SL2SJ0217$-$0513 & HSC R & 1.30 & 1.69 & 4.04 & (0.25, 1.50) & 4/40 \\
J0223$-$0534 & HSC G & 1.27 & 2.82 & 3.70 & (0.25, 1.11) & 27/40 \\
J0226$-$0420 & HSC R & 1.19 & 4.40 & 1.95 & (0.25, 0.96) & 24/40 \\
J1406$+$5226 & CFHT R & 1.19 & 0.63 & 1.68 & (0.28, 1.60) & 8/40 \\
SDSSJ1416$+$5136 & CFHT G & 1.28 & 5.11 & 6.96 & --- & 0/0 \\
\bottomrule
\end{tabular}

\end{table}

\end{appendix}

\end{document}